\documentclass[acmsmall]{acmart}
\copyrightyear{2024}
\acmYear{2024}
\setcopyright{acmlicensed}
\acmConference[DIS '24]{Designing Interactive Systems Conference}{July 1--5, 2024}{IT University of Copenhagen, Denmark}
\acmBooktitle{Designing Interactive Systems Conference (DIS '24), July 1--5, 2024, IT University of Copenhagen, Denmark}
\acmDOI{10.1145/3643834.3661581}
\acmISBN{979-8-4007-0583-0/24/07}

\acmPrice{15.00}
\usepackage{booktabs}
\definecolor{oxfordblue}{rgb}{0.0, 0.13, 0.28}
\definecolor{harvardcrimson}{rgb}{0.79, 0.0, 0.09}
\definecolor{dartmouthgreen}{rgb}{0.05, 0.5, 0.06}
\definecolor{princetonorange}{rgb}{1.0, 0.56, 0.0}
\definecolor{yaleblue}{rgb}{0.06, 0.3, 0.57}
\definecolor{usccardinal}{rgb}{0.6, 0.0, 0.0}
\definecolor{uclablue}{rgb}{0.33, 0.41, 0.58}
\definecolor{msugreen}{rgb}{0.09, 0.27, 0.23}
\definecolor{cornellred}{rgb}{0.7, 0.11, 0.11}
\definecolor{pomegranate}{RGB}{192, 57, 43}
\definecolor{anti-pomegranate}{RGB}{43,178,192}
\definecolor{alizarin}{RGB}{231, 76, 60}
\definecolor{anti-belize}{RGB}{185, 41, 56}
\definecolor{belize}{RGB}{41, 128, 185}
\definecolor{peter}{RGB}{52, 152, 219}
\definecolor{green}{RGB}{22, 160, 133}
\definecolor{anti-green}{RGB}{160,22,118}
\definecolor{turquoise}{RGB}{26, 188, 156}
\definecolor{pumpkin}{RGB}{211, 84, 0}
\definecolor{anti-pumpkin}{RGB}{0,22,211}
\definecolor{carrot}{RGB}{230, 126, 34}
\definecolor{wisteria}{RGB}{142, 68, 173}
\definecolor{anti-wisteria}{RGB}{99,173,68}
\definecolor{amethyst}{RGB}{155, 89, 182}
\definecolor{nephritis}{RGB}{39, 174, 96}
\definecolor{anti-nephritis}{RGB}{174,39,117}

\newcommand{\pzh}[1]{{\color{black} #1}}
\newcommand{\peng}[1]{{\color{black} #1}}
\newcommand{\zhenhui}[1]{{\color{black} #1}}
\newcommand{\xingbo}[1]{{\textcolor{black}{#1}}}
\newcommand{\wxb}[1]{{\textcolor{black}{#1}}}
\newcommand{\chen}[1]{{\color{black} #1}}

\newcommand{\cqyrevise}[1]{{\color{black} #1}}

\newcommand{\qiaoyi}[1]{{\color{black} #1}}
\newcommand{\chenqy}[1]{{\color{black} #1}}

\usepackage{float}
\usepackage{xargs} % Used for new commands with optional arguments
\usepackage{soul}  % Used for custom comments
\usepackage{color} % Used for custom colors in comments
\usepackage{xspace}
\usepackage{listings}
\usepackage{enumitem}
\newcommand{\eg}{{\it e.g.,\ }}
\newcommand{\etal}{{\it et al.\ }}

\newcommand{\ie}{{\it i.e.,\ }}

\definecolor{activegold}{RGB}{255,193,61}

\definecolor{lightorange}{RGB}{230, 170, 50}
\definecolor{lightgreen}{RGB}{121,210,121}
\definecolor{lightteal}{RGB}{121,199,210}
\definecolor{lightblue}{RGB}{100,212,239}
\definecolor{lightpurple}{RGB}{153,102,255}
\definecolor{lightred}{RGB}{245, 132, 120}
\definecolor{red}{RGB}{178,34,34}
\definecolor{gray}{RGB}{166,166,166}

\newcommandx{\guest}[3][1=]
    {\setulcolor{lightorange}{\ul{#1}} \textcolor{lightorange}
    {[\textbf{Peer:} #3]}}
\newcommandx{\cqy}[2][1=]
    {\setulcolor{lightgreen}{\ul{#1}} \textcolor{lightgreen}
    {[\textbf{Qiaoyi:} #2]}}
\newcommandx{\zh}[2][1=]
    {\setulcolor{lightteal}{\ul{#1}} \textcolor{lightteal}
    {[\textbf{Peng:} #2]}}

\newcommand{\name}{{\textit{RetAssist}}}
    
\begin{document}

\title[\name{}]{\name{}: Facilitating Vocabulary Learners with Generative Images in Story Retelling Practices}

\author{Qiaoyi Chen}
\email{chenqy99@mail2.sysu.edu.cn}
\affiliation{%
  \institution{Sun Yat-sen University}
  \city{Zhuhai}
  \country{China}
}

\author{Siyu Liu}
\email{Liusy89@mail2.sysu.edu.cn}
\affiliation{%
  \institution{Sun Yat-sen University}
  \city{Zhuhai}
  \country{China}
}

\author{Kaihui Huang}
\email{huangkh26@mail2.sysu.edu.cn}
\affiliation{%
  \institution{Sun Yat-sen University}
  \city{Zhuhai}
  \country{China}
}

\author{Xingbo Wang}
\email{xiw4011@med.cornell.edu}
\affiliation{%
  \institution{Cornell University}
  \city{New York}
  \country{United States}
}

\author{Xiaojuan Ma}
\email{mxj@cse.ust.hk}
\affiliation{%
  \institution{The Hong Kong University of Science and Technology}
  \city{Hong Kong}
  \country{China}
}

\author{Junkai Zhu}
\email{zhujunkai@hotmail.com}
\affiliation{%
  \institution{Guangdong Polytechnic of Industry and Commerce}
  \city{Guangzhou}
  \country{China}
}

\author{Zhenhui Peng}
\authornote{Corresponding author.}
\email{pengzhh29@mail.sysu.edu.cn}
\affiliation{%
  \institution{Sun Yat-sen University}
  \city{Zhuhai}
  \country{China}
}

\renewcommand{\shortauthors}{Qiaoyi Chen et al.}

\begin{abstract}

\chen{
Reading and repeatedly retelling a short story is a common and effective approach to learning the meanings and usages of target words. However, learners often struggle with comprehending, recalling, and retelling the story contexts of these target words. Inspired by the Cognitive Theory of Multimedia Learning, we propose a computational workflow to generate relevant images paired with stories. Based on the workflow, we work with learners and teachers to iteratively design an interactive vocabulary learning system named \name{}. 
\zhenhui{It can generate sentence-level images of a story to facilitate the understanding and recall of the target words in the story retelling practices.}
\chenqy{
Our within-subjects study (N=24) shows that compared to a baseline system without generative images, \name{} significantly improves learners' fluency in expressing with target words. 
Participants also feel that \name{} eases their learning workload and is more useful. 
}
We discuss insights into leveraging text-to-image generative models to support learning tasks.

% Reading and repeatedly retelling a short story is a common and effective approach to learning the meanings and usages of target words. 
% However, learners often struggle with comprehending, recalling, and retelling the story contexts of these target words. 
% Inspired by the Cognitive Theory of Multimedia Learning, we propose a computational workflow to generate relevant images paired with stories. 
% % to enhance the understanding and recall of the target words and their contextualized usage
% Based on the workflow, we work with learners and teachers to iteratively design an interactive vocabulary learning system named \name{}. 
% It can generate contextually relevant images, facilitating the understanding and recall of the target words and their contextualized usage. 
% % aiding in story comprehension and facilitating repeated retelling exercises. 
% Our within-subjects study (N=24) shows that compared to a baseline system without images, \name{} significantly improves the learning outcome on mastering target words’ meanings and expressions in the story retelling practices. 
% Participants also feel that \name{} eases their learning workload and is significantly more useful. 
% We discuss insights into leveraging text-to-image generative models to support learning tasks. 
}

\end{abstract}


\begin{CCSXML}
<ccs2012>
   <concept>
       <concept_id>10003120.10003121.10003129</concept_id>
       <concept_desc>Human-centered computing~Interactive systems and tools</concept_desc>
       <concept_significance>500</concept_significance>
       </concept>
   <concept>
       <concept_id>10003120.10003121.10011748</concept_id>
       <concept_desc>Human-centered computing~Empirical studies in HCI</concept_desc>
       <concept_significance>300</concept_significance>
       </concept>
 </ccs2012>
\end{CCSXML}

\ccsdesc[500]{Human-centered computing~Interactive systems and tools}
\ccsdesc[300]{Human-centered computing~Empirical studies in HCI}



\keywords{Vocabulary learning, story retelling, image generation}
% vocabulary learning, story retelling, image generation, adaptive assistance}


\received{February 2024}
\received[revised]{April 2024}
\received[accepted]{May 2024}

\maketitle
\section{Introduction}
% 11111
% 1
% \pzh{
% Vocabulary is the foundation of all languages and expanding it is essential for language learning. 
%%%%%%%%%%%%%%%%%%%%%%%%%%%%%%%%%%%%%%%%%%%%%%%%%%%%%%%%%%%%%%%%%%%%%%%%%%%%%%%%%%%%%%%%%%%%%%%%%%%%%%%%%%%%%%%%%%%%%%%%%%%%%%%%%%%%%%%%
% [[vocabulary learning practice]]
\peng{
Learning vocabulary in meaningful contexts, such as stories and images in language learning textbooks, and video clips from movies, is a common and effective practice 
% \qiaoyi{for English-as-the-Second-Language learners} 
as it enables deep and active processing of vocabulary (\eg word associations, logic) \cite{oxford1994second}.
% Learning vocabulary is the basic practice for anyone who wants to master a language. 
% Educators and 
Human-Computer Interaction (HCI) researchers have explored various technologies to support vocabulary learners with meaningful contexts in various learning activities, \eg \textit{ViVo} in watching videos \cite{10.1145/3025453.3025779}, \textit{VocabEncounter} in reading online articles \cite{arakawa2022vocabencounter}, and \textit{EnglishBot} in conversing with others \cite{ruan2021englishbot}. 
In this paper, we focus on the story retelling activity that \chenqy{encourages English-as-the-Second-Language vocabulary learners} to integrate, reconstruct, and demonstrate the contextualized use of the target words in a short story \cite{morrow1985retelling, merritt1989narrative, gibson2003power,kutuk2007effect}. 
% , and it is also a common vocabulary learning measure which is included in the College Entrance Examination in China. 
This practice typically involves two stages -- story comprehension and repeated retelling \cite{nguyen2019effect}, \ie the learner first reads or listens to a short story that contains a set of target words to comprehend its main idea and then verbally retells it for multiple rounds. %, \eg to the listeners if any. 
% When applied to vocabulary learning, the used short story would contain a set of target words that are required to learn \cite{kutuk2007effect}. 
% [[what are the advantages of it (comprehensive and expressive)? ]]
% Through story retelling, learners can understand the meanings of target words by reading a coherent story context, which is a typical meaning-focused input vocabulary learning activity \cite{izumi2002output,nation2007four}. 
% They can also practice the usage of these words to develop their fluency in language expression \cite{nation2007four}.
Several studies on language education have demonstrated the effectiveness of story retelling for vocabulary learning \cite{dunst2012children, gibson2003power, miller2008power}, especially in remembering the meanings of target words and using them in verbal expressions \cite{dunst2012children, ruan2021englishbot}. 
\chen{In fact, story retelling has been included in the English test of the College Entrance Examination in China \footnote{https://gaokao.eol.cn/guang\_dong/dongtai/201811/t20181101\_1631228.shtml}. }
% [refs]. 
}

% [[what are the challenges of it (images may deal with the challenges)? ]]
\peng{
However, the story retelling practice is often challenging \qiaoyi{for learners of second language vocabulary. }
% for vocabulary learners. 
%\zh{To do: mention the two main challanges below.}
% To explore the challenges that learners may face in story retelling and the effective practices used by the pedagogues to address these challenges, we referenced work related to vocabulary learning and conducted a formative study with three English teachers (E1-E3) and seven English learners (S1-S7).
% our formative study with xxx participants reveal that ...
% During the story comprehension stage, learners focus on the contextualised use of target vocabulary in the story text which they are unfamiliar with and pay close attention to the details, making inferences and relating the contextual meaning of the vocabulary in the story to the possible word meanings of the target pairs previously skimmed \cite{nation2007four, morrow1985retelling, kintsch2005comprehension}.
For one thing, in the story comprehension stage, learners need to associate the meanings of target words with the story context and memorize the story flow for the later repeated retelling practice \cite{nation2007four, morrow1985retelling, kintsch2005comprehension}. 
For another, in the repeated retelling stage, they should repeatedly narrate the read story with requirements on the correct usage of target words and fluency in speaking them out in the story \cite{nation2007four, morrow1985retelling, kintsch2005comprehension}. 
In other words, it requires learners to 
% It is a non-trivial practice that requires learners' efforts on 
understand, memorize, recall, organize, and speak the target words and associated story \cite{kintsch2005comprehension}. 
This becomes more challenging when there is a time limit for each round of repeated retelling, which could help to develop language fluency under pressure \cite{nation2007four}.
% This becomes more challenging when there is a time limit for each story retelling session, which is a common trick as it could help to develop language fluency \cite{nation2007four}. 
}
\peng{
% In traditional classroom settings, the learners can get assistance from teachers in their story retelling practices. 
% For example, teachers can prepare relevant images for the stories from course books or online resources. 
%%%%%%%%%%%%%%%%%%%%%%%%%%%%%%%%%%%%%%%%%%%%%%%%%%%%%%%%%%%%%%%%%%%%%%%%%%%%%%%%%%%%%%%%%%%%%%%%%%%%%%%%%%%%%%%%%%%%%%%%%%%%%%%%%%%%%%%%
% [[Image argument]]
Images related to the story can help vocabulary learners cope with these two challenges during the story retelling practice. %practical material to
% [[Theory]]
As suggested by the Cognitive Theory of Multimedia Learning (CTML) \cite{paivio1980dual}, building mental representations from text and visual elements could facilitate comprehension and recall of words and their contextualized usage \cite{oktarina2020effectiveness, filippatou1996pictures, mayer2002multimedia, sorden2012cognitive}. 
\chen{
In the context of second language acquisition, individuals tend to subvocally articulate the text associated with visual stimuli in their native language \cite{paivio2014bilingual}.  
% \lsyrevise{
% In the context of second language acquisition, individuals tend to organize visual stimuli with their native language while memorizing textual stories \cite{paivio2014bilingual}.
% In the context of second language acquisition, individuals instinctively use their native language to memorize and organize storylines with visual stimuli \cite{paivio2014bilingual}.
% }
Thus, compared to learning without visual aids, non-verbal modalities such as images bridge the gap between two different languages, \pzh{which would enhance} the likelihood of recalling the second language \cite{paivio1990mental, paivio2014bilingual}. 
% Specifically, learners covertly pronounce the corresponding text of the images in their native language. \cite{paivio2014bilingual}
% As a non-verbal modality, images improve L2 learning by linking two languages \cite{paivio2014bilingual, paivio1990mental}.
% The text in the native language and the visual elements converge on the foreign language responses, which increases the probability of recall relative to the condition without visual aids \cite{paivio2014bilingual}.
}
% [[Example Using Image]]
Given these benefits, language educators widely prepare relevant images for the textual stories in course books or online resources, and HCI researchers have proposed vocabulary learning support systems in activities that involve visual elements 
% Prior HCI researchers have explored technologies to support language learners with visual materials 
\cite{10.1145/3025453.3025779, 10.1145/3369824, 10.1145/3462204.3481750}. 
% For instance, \textit{ViVo} \cite{10.1145/3025453.3025779} generates video clips that highlight the target word in the subtitles, and
For example, \textit{CoSpeak} \cite{10.1145/3462204.3481750} uses voice recognition techniques to support students to collaboratively and verbally create a story given an image prompt. % pairs students together to create a story based on an image prompt. 
% [[Traditional]]
% In traditional classroom settings, teachers can prepare relevant images for the stories from course books or online resources. 
\cqyrevise{
However, it is time-consuming and often unavailable to prepare relevant images for the story with any set of target words that users wish to learn in the story retelling practices,  
% Also, inappropriate or irrelevant images tend to confuse learners, causing delays in learning or hindering memorization \cite{hasnine2017algorithm}.
\zhenhui{while irrelevant images would confuse learners and reduce vocabulary learning outcome \cite{hasnine2017algorithm}.} 
% However, it is time-consuming and often unavailable to prepare relevant images for the story with any set of target words that users wish to learn in the story retelling practices. 
% Despite the benefits of images, they are usually unavailable when individuals want to learn a set of customizable target words through story retelling practices at any time. 
}

% It would take a lot of time to curate relevant vivid images, if existed, for any story content. 
% [[Others]]
% According to the Theory of Scaffolding \cite{vygotsky1978mind}, 
% such visual guidance is valuable for students to recall the missed details and can increase students' likelihood of mastering the target words' meanings and expressions \cite{sekeres2016recovering}. 
% help students comprehend and memorize 
% the target words and stories 
% Teachers can also provide timely prompts (\eg hints on the next sentence when students get stuck) during the retelling practice and feedback (\eg whether the target words are correctly used) on students' retelling performance.
% time limit
% Also, there is usually a time limit for each story retelling session, which is a common trick as it could help to develop language fluency \cite{nation2007four, boers2014reappraisal}. 
% This becomes more challenging when there is a time limit for each story retelling session, which is a common trick as it could help to develop language fluency \cite{nation2007four}. 
% \zh{Check https://www.indeed.com/career-advice/career-development/vygotsky-scaffolding You need to describe this theory briefly in Related work} 
% As the learners become more comfortable with the retelling materials and practices, the assistance from the educators lessens until the learner can finish the practice on their own \cite{vygotsky1978mind}. 
%\eg mastering the meanings of target words and being able to  
}

\peng{
%%%%%%%%%%%%%%%%%%%%%%%%%%%%%%%%%%%%%%%%%%%%%%%%%%%%%%%%%%%%%%%%%%%%%%%%%%%%%%%%%%%%%%%%%%%%%%%%%%%%%%%%%%%%%%%%%%%%%%%%%%%%%%%%%%%%%%%%
% [[images dealing with challenges]]
In this work, we explore the design and usage of generative images to facilitate the learning of any target word set via reading and repeatedly retelling a short story that contains these words. 
% customize images from stories containing target vocabulary to aid learners' comprehension and recall of the story.
Our focus is motivated by the benefits of images for vocabulary learning as described above and recent advances in text-to-image generative techniques. 
For example, the pre-trained 
% a certain trained 
Latent Diffusion Model (LDM) \cite{Rombach_2022_CVPR} is able to generate high-quality and content-relevant images given a text prompt. % from text prompts.
% Despite the benefits of visual materials, they are usually unavailable when individuals want to learn a set of customizable target words through story retelling practices at any time. 
% It would take a lot of time to curate relevant vivid images, if existed, for any story content. 
% Recent advances in text-to-image generation techniques \cite{Rombach_2022_CVPR} offer the potentials to tackle this issue. 
% Yet, it is unknown whether the generative images of given textual stories are meaningful aids for supporting story retelling practices. 
% Prior HCI researchers have explored technologies to support language learners with visual materials \cite{10.1145/3025453.3025779, 10.1145/3369824, 10.1145/3462204.3481750}. 
% For instance, \textit{ViVo} \cite{10.1145/3025453.3025779} generates video clips that highlight the target word in the subtitles, and \textit{CoSpeak} \cite{10.1145/3462204.3481750} pairs students together to create a story based on an image prompt. 
% [[generative images are used in ...]]
\chen{
These generative techniques have been used to support the creations of artworks \cite{gatys2015neural}, medical images \cite{frid2018synthetic}, and game characters \cite{emekligil2022game}. 
}
% xxx [refs], and xxx [refs].  
% Generative images have been generally used by artists to create unique and novel artworks and designs \cite{gatys2015neural}.
Nevertheless, little work, if any, has explored generative images for supporting vocabulary learning in the story retelling practices where users should master target words' meanings and verbal expressions. 
% whether the generative images of given textual stories can assist and facilitate vocabulary learners during their story retelling practices. 
Questions arise such as 
1) whether and how text-to-image generative techniques can generate relevant images of any story that covers a target word set,
2) if so, what kinds of support that the generative images can offer in the story retelling practices, 
and 
3) how would the support from generative images impact the users’ vocabulary learning outcome and experience.
% Second, adaptive prompts and feedback from qualified teachers are not always there for students, especially in off-school learning scenarios.
% Prior HCI researchers have explored technologies to replace teachers with peer listeners \cite{10.1145/3462204.3481750} or a conversational agent \cite{ruan2021englishbot}. 
% For example, Ruan \etal proposed a \textit{EnglishBot} that converses with students interactively on college-related topics and provides adaptive feedback for spoken language learning \cite{ruan2021englishbot}.  
% % Nevertheless, their bot targets the listen-and-repeat practices on conversations that are mainly for language learners to practice speaking skills. \zh{Umm... any significant difference from the EnglishBot work?}
% % Focus on enhancing rich verbal expressions in oral communication
% Nevertheless, their bot requires manual effort in preparing learning materials and targets at practicing speaking. 
% % dialoguing activities that are mainly for improving learners' spoken language in conversation. 
% Little work has explored how visual materials can facilitate learners during their story retelling practices. 
% Few works have explored the design, effectiveness, and user experience of  technological adaptive support for story retelling practices with a focus on vocabulary learning. 
}

\peng{
To this end, we seek to provide insights into these questions by designing, developing, and evaluating an intelligent system prototype, \name{}, that can generate relevant images for learning vocabulary in the story retelling practices. %facilitate vocabulary learners in their story retelling practices with generative images. 
% To this end, we build an intelligent system prototype \name{} to facilitate vocabulary learners in their story retelling practices with generative images. 
Here, we target English-as-the-Second-Language (ESL) Chinese learners, \eg high-school or university students in China. 
We take an iterative design approach with insights from educational literature and the involvement of ESL learners and English teachers in this process. 
\chen{
We first develop a text-to-image computational workflow and validate its capability in generating a series of coherent and relevant sentence-level images given any short textual story that contains IELTS \footnote{Short for International English Language Testing System, a globally recognized standardized test designed to assess the English language proficiency of individuals.} target words.  
}
% IELTS \footnote{Short for ..., a ...} target words.  
We then conduct an interview study with seven ESL learners to understand their challenges and needs in the story retelling practices and ask for their comments on the generative images. 
% image prompts in story retell practice by text-to-image generation techniques, and evaluate its feasibility and support for story-retelling-based vocabulary learning.  
% We then conduct an interview study with seven ESL learners to explore their possible challenges and needs in story retelling practices and to ask for their comments on the generative image prompts.  
% and derive design requirements for our system. 
Based on the insights from the interviews and educational literature, we develop a \name{} prototype and seek feedback from another 18 ESL learners and two English teachers to refine it. 
\zhenhui{In the story retelling practice with the refined \name{}, users can first }
% In the story comprehension stage, users can 
read and listen to the story with generative images aligned to each sentence. 
\peng{
% Then, during each round of repeated retelling, users can retell the story with the assistance of images and speech transcript.  
Then, during each round of repeated retelling, users can retell the story by viewing the images.  
% click the image to xxx. 
}
After each round, users can review their performance in the expressions of target words and re-read the story with images. 
% three types of assistance during story retelling practice: 1) generative images in both stages of story comprehension and repeated retelling practice, 2) cloze sentence triggered by learners' retelling stuckness, 3) feedback reflecting semantic correctness of vocabulary use. 
% We seek feedback on \name{}’s design and refine it via a user study with 18 ESL learners and a co-design workshop with two experienced English teachers.
}

\peng{
We conduct a within-subjects study with 24 ESL vocabulary learners to evaluate the impact of \name{}'s function on the generative image on the vocabulary learning outcome and experience. 
% effectiveness and user experience against a baseline tool without the generative images. 
% [Short description about the content of the study]
% Our research questions are: 
% \textbf{RQ1.} How would \name{} affect users’ a) engagement and enjoyment and b) retelling behaviors in their vocabulary-focused story retelling practices?
% \textbf{RQ2.} How would \name{} affect users’ learning outcome regarding the retention and verbal expression of target words in their vocabulary-focused story retelling practices?
% \textbf{RQ3.} How would users perceive \name{} that provides generative images and adaptive assistance in their vocabulary-focused story retelling practices?
% To explore the impact of \name{} to users, we evaluate the learning process, learning outcomes, and users' perceptions of using the system in two experimental conditions.
% [Main findings of the study] 
\chenqy{
% The results show that compared with the baseline system without generative images, participants using \name{} significantly outperform in recalling the target words' meanings and using these words in verbal expressions in the posttest one week after the learning sessions.
% Participants favor the generative images of \name{} for reducing learning workload and aiding recall of the contextual usage of target words in the story. 
The results show that compared with the baseline system without generative images, participants using \name{} significantly outperform in fluently using the target words in verbal expressions.
% in the posttest one week after the learning sessions.
Participants favor the generative images of \name{} for reducing learning workload and aiding recall of the contextual usage of target words in the story. 
}
% As for learning experience, learning with \name{} is significantly effective for easing users' learning workload and is perceived as more useful. 
% helps learners better master the target words' verbal expressions, especially on their pronunciations and correct usage in story context. 
% With either system, participants master the target words' meanings well after the story retelling practices. 
% Participants report that learning with \name{} is significantly more engaging and enjoyable than practicing with the baseline tool. 
Based on our findings, we highlight the value of text-to-image generative techniques in offering useful learning materials and enjoyable learning experiences. 
We further discuss design considerations for future vocabulary learning support systems \zhenhui{and the impact of our work on generative AIs for education}. % improving story-retelling-based vocabulary learning systems. 
% It demonstrates that \name{} helps learners master the target words' expressions more effectively than Read-and-Retell without our proposed features and also improves users' learning experience and satisfaction.
}

\peng{
Our work makes three contributions. 
First, we present a vocabulary learning system \name{} that uses generative images to facilitate users to master target words' meanings and expressions via story retelling practices. 
Second, our design and evaluation of \name{} provide first-hand findings on the feasibility, effectiveness, and user experience of applying text-to-image generative models to vocabulary learning. 
Third, we propose a story text-to-image generation workflow and offer design considerations of leveraging generative models to support learning tasks. 
}

% The major contributions of our work are summarized as follows:
% \begin{itemize}
% \item[$\bullet$] An interactive system \name{} that facilitates vocabulary learners with generative images to master the target words' meanings and verbal expressions through story retelling practices. %  and develop fluency in their usage 
% %enables users to learn the contextualized use of target English words by understanding and repeatedly retelling meaningful stories with intelligent support such as visual aids and prompts for contextualized target words.
% % supported by a series of generative images and specific eliciting questions.
% \item[$\bullet$] A computational workflow that generates images relevant to any textual story to foster comprehension and recall of the story.
% % that can assist EFL learners in story comprehension, integration and reconstruction.
% \item[$\bullet$] Empirical findings from a within-subjects user study that demonstrate the learning outcome and experience of \name{}.
% % improving the retention rates and speaking fluency of target words.  
% %discussion that compares the effectiveness of vocabulary learning in both mode with visual information and prompts support and without these support.
% \item[$\bullet$] Design considerations for improving the effectiveness and user experience of future vocabulary learning support tools. 
% \end{itemize}

\section{Related Work}
% \pzh{Reminder: at the end of each subsection, you should describe the link and difference between your work and related works.}
% \subsection{Story Retelling for Vocabulary Learning}
% \subsection{Vocabulary Learning Support Systems}
% \subsection{Text-to-Image Generation / or Story Visualization?}
\peng{
We introduce prior studies that motivate, inspire, and support the design of \name{}, including story retelling for vocabulary learning, vocabulary learning systems, and text-to-image generation techniques.
}

\subsection{Story Retelling for Vocabulary Learning}
% 1) Comparison with other story-based approaches to vocabulary learning \cite{ghorbani2014story} 
% 2) The two stages of story retelling \cite{nguyen2019effect} 
% 3) the aids that the pedagogue can provide during the corresponding stages \cite{dunst2012children}

% \textbf{2) The process of the story retelling} consists of both story comprehension and retelling practice \cite{nguyen2019effect}.

% 11111
% 1
% Story retelling is a common teaching method used in vocabulary learning, as it encourages learners to integrate, reconstruct and demonstrate the contextual use of vocabulary \cite{morrow1985retelling, merritt1989narrative, gibson2003power} so that enables learners to memorize the explanation of target words and enhance their capacity to develop appropriate language expressions using the words \cite{ghorbani2014story, dunst2012children}. 
\peng{
Story retelling is a well-recognized approach that helps students acquire vocabulary and skills like reading, listening, and speaking in language learning and teaching \cite{nguyen2019effect,nation2007four, morrow1985retelling, merritt1989narrative}. 
% Additionally, the prevalence of this vocabulary learning measure is exemplified through its inclusion in the College Entrance Examination in China.
% Story retelling is a common teaching method used in vocabulary learning. 
% It encourages learners to integrate, reconstruct and demonstrate the contextual use of vocabulary \cite{morrow1985retelling, merritt1989narrative, gibson2003power}, enabling them to memorize the explanation of target words and enhance their capacity to develop appropriate language expressions using the words \cite{ghorbani2014story, dunst2012children}.
% \zh{General writing guidance: a long sentence (\eg longer than three lines) should be divided.}
% 2
A story retelling practice normally consists of two stages, \ie story comprehension in which learners listen or read a given story, and repeated retelling in which they speak it out for several times within a time limit \cite{nguyen2019effect, nation2007four}. 
% It is a well-recognized approach for practicing language skills like reading, listening, and speaking \cite{nguyen2019effect,ruan2021englishbot}. 
% Practicing story retelling includes two stages of story comprehension and repeated retelling practice \cite{nguyen2019effect}.
% In this work, we focus on the story retelling approach for vocabulary learning, which 
% The story retelling approach has been shown effective in traditional language courses \cite{nation2007four, morrow1985retelling, merritt1989narrative}. 
% 3
As suggested by Nation \etal \cite{nation2007four}, it is a practice that properly integrates four typical strands of activities, \ie meaning-focused input, meaning-focused output, language-focused learning, and fluency development. 
% Nation \cite{nation2007four} classified activities in a language course into the four strands of meaning-focused input, meaning-focused output, language-focused learning, and fluency development.
% Analyzing the characteristics of these four strands, the two stages are considered to be meaning-focused input and fluency development of output.
% \zh{Should briefly introduce these two strands.}
% like the contextualized use of vocabulary in learning materials 
% Analyzing the characteristics of Nation's four strands theory, these two stages correspond respectively to meaning-focused input and fluency development with productive learning features \cite{nation2007four}. 
First, in the story comprehension stage, learners focus on understanding the given story with target words -- using language receptively (meaning-focus input)  \cite{nation2007four}. 
% Meaning-focused input learning strand is to acquire some knowledge about unknown language items through context clues and background knowledge \cite{nation2007four}. 
Next, in the repeated retelling stage, learners are required to correctly and fluently speak the story out -- using language productively (meaning-focus output) \cite{nation2007four}. 
Moreover, in the whole practice, learners should specifically pay attention to the meanings, pronunciations, and correct usages of the target words -- deliberate learning of language features (language-focused learning) \cite{nation2007four}. 
Lastly, the practice requires learners to make the best use of what they already know to perform well in retelling under time pressure \cite{morrow1985retelling, kintsch2005comprehension,izumi2002output, boers2014reappraisal}, which is a typical fluency development learning activity \cite{nation2007four}. 
In all, story retelling encourages learners to integrate, reconstruct, and demonstrate the contextual use of vocabulary \cite{morrow1985retelling, merritt1989narrative, gibson2003power}. 
The expected learning outcome is, therefore, not only on memorization of target words' meanings but also on the capacity of using the words correctly and fluently in language expressions \cite{ghorbani2014story, dunst2012children}.
}
\peng{
Given these requirements of reading, interpreting, memorizing, and speaking the story with target words \cite{nation2007four, morrow1985retelling, kintsch2005comprehension}, story retelling is often challenging for learners. 
% The story retelling practice is challenging.
% Learners will encounter some challenges in both stages of story comprehension and repeated retelling, and assistance corresponding to these challenges is considered to influence the effectiveness of practicing story retelling \cite{dunst2012children}.
% However, learners at both stages will face a number of challenges that can affect the effectiveness of the activity. 
% 2
% During the story comprehension stage, learners need to correlate the meaning of the target words with the context of the story and remember the flow of the story for later retelling practice \cite{nation2007four, morrow1985retelling, kintsch2005comprehension}.
% In the stage of story comprehension, learners need to have a good understanding of the story content for being able to retell a story effectively \cite{kintsch2005comprehension} and the aid of visual elements (such as images) can help learners to understand the story correctly \cite{oktarina2020effectiveness, filippatou1996pictures}. 
% This may involve using visualization techniques to help learners understand the story \cite{dunst2012children, willis2007doing}.
% 3
% In the repeated retelling stage, the narrative is repeated for the read story, requiring accurate use of the target words and fluency in the story \cite{nation2007four, morrow1985retelling, kintsch2005comprehension}. 
% The time limit is gradually reduced for each retelling, a common technique as it helps to improve verbal fluency, which also makes the task more challenging \cite{nation2007four}.
\chen{Traditionally, there are additional materials (\eg images and props) to the textual story and in-situ guidance from teachers (\eg prompting phrases) to assist learners in the story retelling practices \cite{dunst2012children, douglas2001teaching}. }
% some intervention during this practice assists vocabulary learners \cite{dunst2012children, douglas2001teaching}. 
\zhenhui{
Images relevant to the story, for example, are beneficial in that they can help learners remember the words' meanings and comprehend the story \cite{oktarina2020effectiveness, filippatou1996pictures, alhamami2016vocabulary}. 
Images can also serve as the visual guidance that helps learners recall the story and target words when they get stuck in the repeated retelling stage \cite{vygotsky1978mind}. 
% Broadly speaking, images are highly recommended by the Cognitive Theory of Multimedia Learning (CTML)  \cite{paivio1980dual, oktarina2020effectiveness, filippatou1996pictures, mayer2002multimedia, sorden2012cognitive} and the Bilingual Dual Coding Theory (BDCT) \cite{paivio2014bilingual, paivio1990mental} in language learning, which argues that learners can recall textual information (\eg the story in our case) based on relevant visual information.
}
\qiaoyi{
According to Cognitive Theory of Multimedia Learning (CTML) \cite{oktarina2020effectiveness, filippatou1996pictures, mayer2002multimedia, sorden2012cognitive} and Dual Coding Theory (DCT) \cite{paivio1980dual, paivio1990mental}, 
building mental representations from text and visual elements could enhance the encoding and retention of information by leveraging dual coding, which taps into both verbal and visual processing systems in the brain.
As the extension of DCT, Bilingual Dual Coding Theory (BDCT) \cite{paivio2014bilingual} suggests that images enhance second language learning since learners covertly pronounce the content of the images in their native language and the content and the images converge on the foreign language responses, increasing the probability of recall relative to the condition without images.
Furthermore, Mayer identified the twelve multimedia instructional principles to address the issue of how to structure multimedia instructional practices and employ more effective cognitive strategies to help people learn efficiently \cite{mayer2002multimedia}. For instance, the spatial contiguity principle \cite{mayer2002multimedia} indicates that people learn better when corresponding words and images are placed near each other rather than far from each other on the page or screen.
% Broadly speaking, images are recommended to be included in language learning in an appropriate way, which argues that learners can recall textual information (e.g., the story in our case) based on relevant visual information.
}
\chenqy{
In summary, these principles suggest that relevant visuals (\eg images) can significantly aid in recalling textual information (\eg story in our case) of vocabulary to facilitate vocabulary learning. 
Despite the clear benefits, selecting appropriate images that align closely with the textual content remains a considerable challenge \cite{hasnine2018image}. 
}
\zhenhui{
Our work is motivated by the benefits of story retelling practices for enhancing understanding and expression of target words and the helpfulness of images for assisting users in these practices. 
% Our work is motivated by the benefit of story retelling practices for learners to enhance understanding of target vocabulary and expression using target vocabulary \cite{dunst2012children}.
Instead of requiring human effort to prepare the images, we propose to generate relevant images to any story that covers the target words that users wish to learn. %to use machine-generative images as visual aids for
}
}

\subsection{Vocabulary Learning Systems}
\peng{
Existing HCI researchers have explored various intelligent systems to support vocabulary learning. 
Broadly speaking, they are either based on word lists or meaningful contexts. 
The former type of vocabulary learning system aims to facilitate quick memorization of target words in a list.  
Previous work has incorporated models of users' memory cycles and individualized learning styles into these systems, such that they can recommend a set of target words with appropriate levels of difficulty and repetition frequency \cite{nioche2021improving, zeng2011interactive, chen2008personalized}.
For example, Chen \etal \cite{chen2008personalized} proposed a personalized mobile English vocabulary learning system based on Item Response Theory and the learning memory cycle. 
% \cqyrevise{
% \zhenhui{[Peng: Confusing. Contextual images could also be the context of the image. Consider moving this example in the second category.]}
% As for learning target words with images, \textit{AIVAS} \cite{hasnine2017algorithm}, supported by an image reranking algorithm, determines the most appropriate image to represent a concrete word. 
% However, since the words are not in context, it has limitations in dealing with the words with polysemy and some words are difficult to express with contextual images. 
% }

Context-based vocabulary learning systems leverage various forms of materials such as stories \cite{10.1145/3462204.3481750}, videos \cite{10.1145/3025453.3025779}, and online articles \cite{arakawa2022vocabencounter} to help users learn vocabulary. % in context. 
\cqyrevise{
For example, 
\xingbo{\textit{VocabEncounter} \cite{arakawa2022vocabencounter} encapsulates target vocabulary into the context of online \zhenhui{articles}, while \textit{ARLang} \cite{caetano2023arlang} visualizes bilingual labels on physical objects outdoors in AR environment to support the micro-learning of language within its spatial context.
Additionally, \textit{EnglishBot} \cite{ruan2021englishbot}, a language learning chatbot, engages students in interactive conversations on college-related topics to learn English.
\qiaoyi{
Learners can click as needed to receive answer prompts provided in their native language, ensuring smooth conversations with \textit{EnglishBot} \cite{ruan2021englishbot}. 
In line with \textit{EnglishBot}'s method, \name{} allows users to autonomously click on corresponding images based on their current progress when comprehending or retelling stories. 
}
Some studies use images as visual contexts to support vocabulary learning. \textit{AIVAS} \cite{hasnine2017algorithm} uses an image reranking algorithm to select images that prominently contain relevant objects in the middle ground, thus aiding in representing concrete nouns effectively. Furthermore, \textit{FCAI} \cite{hasnine2018image} considers users' personal information, learning time, and location to recommend contextually appropriate images that best represent the target words.
\qiaoyi{
Both \textit{AIVAS} and \textit{FCAI} focus on searching for appropriate images for target words, whereas the images in \name{} need to represent the story content, potentially favoring a generative approach.
}
}
}
% [main results]
\xingbo{Story retelling also facilitates contextualized vocabulary learning.
}
% As for story retelling support, 
\textit{CoSpeak} \cite{10.1145/3462204.3481750} provides an application for learners to practice speaking English by pairing them together to co-create a story with an image prompt based on the ongoing topic in class. 
\qiaoyi{
% The evaluation study shows that \textit{CoSpeak} is helpful in improving users' oral expression in English \cite{10.1145/3462204.3481750}. 
Unlike the focus of our study on individual learners using \name{} for vocabulary learning through story retelling, \textit{CoSpeak} concentrates on enhancing English oral expression through dialogues between two individuals in thematic story settings.
}
% It prepares the image prompts by English teachers and gets feedback on their dialogue from senior students \cite{10.1145/3462204.3481750}.
% Practice speaking English in a conversation based on an ongoing topic in class
% \zh{How its study show?}
% This type of speaking practice, which reproduces a conversation scene, 
% It is shown that this application is helpful in improving users' spoken English expressions. 
% The teacher and the students found the application helpful to learn spoken English in daily conversations and increase English speaking confidence in a creative and fun way.
% Creating a story is an online collaborative activity through which a student can co-create an audio story with a peer. Students are paired together in class by the teacher to create a story. Image prompts consisting of different characters in a situation are provided based on the ongoing topic in class. Each student is assigned a character. To aid the student to start speaking, the beginning of the audio story and an audio prompt to continue the story is provided. 

Our proposed \name{} falls into the category of context-based vocabulary learning systems. 
% Unlike \textit{CoSpeak} preparing conversation aids by English teachers and getting feedback from senior students, 
% and \textit{EnglishBot} supplying evaluative feedback and prompts in the user’s native language for conversation exercise, 
\qiaoyi{
% In contrast to many systems (\eg \textit{VocabEncounter}, \textit{EnglishBot}, and \textit{CoSpeak}) that largely leverage existing materials as contexts, \name{} generates images relevant to the story as contexts to support vocabulary learners in the story retelling practices. 
\name{} not only integrates vocabulary into the story to provide textual context, but also generates a set of related images for the story that serve as the visual context to help individual vocabulary learners acquire vocabulary through story retelling.
}
}

\subsection{Text-to-Image Generation Techniques} %Image Generation for Story
\peng{
\wxb{As suggested by \qiaoyi{the Dual Coding Theory \cite{paivio1980dual, paivio1990mental}}, textual stories paired with relevant images can facilitate vocabulary learning.}
Recent advances in text-to-image generative techniques offer great potential for preparing visual aids for any story that covers learners' interested words. 
\chen{
Text-to-image generative models normally take a text prompt as input and output one or multiple images that are related to the text content. 
One of the early representatives is Diffusion Probabilistic Model (DM) \cite{sohl2015deep}, which achieved state-of-the-art results in density estimation (\ie how well the model captures the probability distribution of the dataset) \cite{kingma2021variational} as well as in sample quality (\ie how well the model generates data samples that closely resemble real data from that distribution) \cite{dhariwal2021diffusion}.
}
% Density estimation is the process of modeling the probability distribution of a dataset, while sample quality measures how well a generative model can produce data samples that resemble real data. The Diffusion Probabilistic Model (DM) excels in both tasks, making it effective in generating high-quality data samples and capturing complex data distributions.
% explain: the result of research studies that evaluated the performance of different models in terms of both density estimation and sample quality
DMs use deep learning techniques to generate high-quality images from text prompts but have the downside of low inference speed. 
% which achieved the state-of-the-art results in density estimation \cite{kingma2021variational} (A statistical method used to estimate the probability density function of a random variable based on a set of data)
% % \zh{what's this} 
% \zh{of generative image?}
% as well as in sample quality \zh{of generative image?} \cite{dhariwal2021diffusion}. 
% However, DMs have the downside of low inference speed. 
To address the drawback, recent approaches widely leverage Latent Diffusion Models (LDMs) \cite{Rombach_2022_CVPR}, which work on a compressed latent space of lower dimensionality and speed up inference with almost no reduction in image synthesis quality. 
\wxb{In this work, we use a state-of-the-art LDM for generating images from text prompts.}
% In this work, we use a certain trained LDM for generating images from text prompts. 
}

Compared to the traditional text-to-image tasks that generate images from a text prompt, image sequence generation for stories is more challenging as it needs to generate a sequence of coherent and consistent images for a story that contains multiple sentences. 
To enhance the image quality and the consistency of the generated sequences, \textit{StoryGAN} consists of a deep Context Encoder that dynamically tracks the story flow, and two discriminators at the story and image levels \cite{li2019storygan}.
\zhenhui{
\textit {Neural Storyboard Artist} visualizes the story in the form of a comic strip through the retrieval of multiple related images from the story content and several image rendering steps like segmenting relevant regions and converting the images into cartoon style \cite{chen2019neural}. %style unification 
% To unify the styles of retrieved image sequences, Chen \etal convert the images into cartoon style, which ensures being textual relevant and visually consistent \cite{chen2019neural}.
% \textit {Neural Storyboard Artist} visualizes the story in the form of a comic strip through the retrieval of multiple related images from the story content and some image rendering steps like relevant region segmentation and style unification \cite{chen2019neural}.
}
Nevertheless, previous story image generation techniques prioritize \wxb{coherence of the whole story flow, which may overlook the contexts (\eg sentences) containing target words for learning. }
% As suggested by the Bilingual Dual Coding Theory \cite{paivio1980dual}, images should semantically match with the text content so as to boost language learning. 
% general storylines, 
% which may not be able to reveal the content of each sentence that contains target words. 
% In order to provide images for story retelling practices, we need to consider a computational workflow based on text-to-image generation techniques.
% As a consequence, 
To facilitate vocabulary learning based on story retelling practices, 
\wxb{we segment the whole story into sentences to provide rich contexts for target words. 
The sentences are used as prompts to generate a sequence of relevant images. 
Specifically, we use a Stable Diffusion model~\cite{runwayml:stable-diffusion-v1-5} to convert sentences into images and apply a cross-modal model, CLIP~\cite{radford2021learning}, to select the most relevant one for each sentence.
\zhenhui{To improve the visual consistency and coherence of the image sequence, we follow \cite{chen2019neural} and use a style transfer model \cite{chen2018cartoongan} to unify the images with cartoon styles.}
% \cqyrevise{
% Since the sequence of images related to the story should look visually consistent with coherent style \cite{chen2019neural}, we use a cartoon style transfer model \cite{chen2018cartoongan} to unify the image styles. 
% To enhance consistencies between images~\cite{chen2019neural}, we use a cartoon style transfer model~\cite{chen2018cartoongan} to unify the image styles.}
% }
% we weakly consider the coherence between images and directly propose each sentence as a prompt for generating one of the images in the image sequence. 
% Also, we utilize a pretrained cross-modal model, CLIP \cite{radford2021learning}, to select the most relevant images from the multiple images generated for each sentence and unify the image styles using a cartoon style transfer model \cite{chen2018cartoongan} so as to mitigate some inconsistencies between images \cite{chen2019neural}. 
\wxb{Thereafter, our proposed computational workflow can generate relevant and style-consistent images for story sentences as meaningful contexts to facilitate story retelling for vocabulary learning.}
}
\section{Design Process}
% \section{Feasibility and Design Requirements of Generative Images for Vocabulary Learning via Story Retelling}

\begin{figure}[htbp]
    \centering
    \includegraphics[width=11cm]{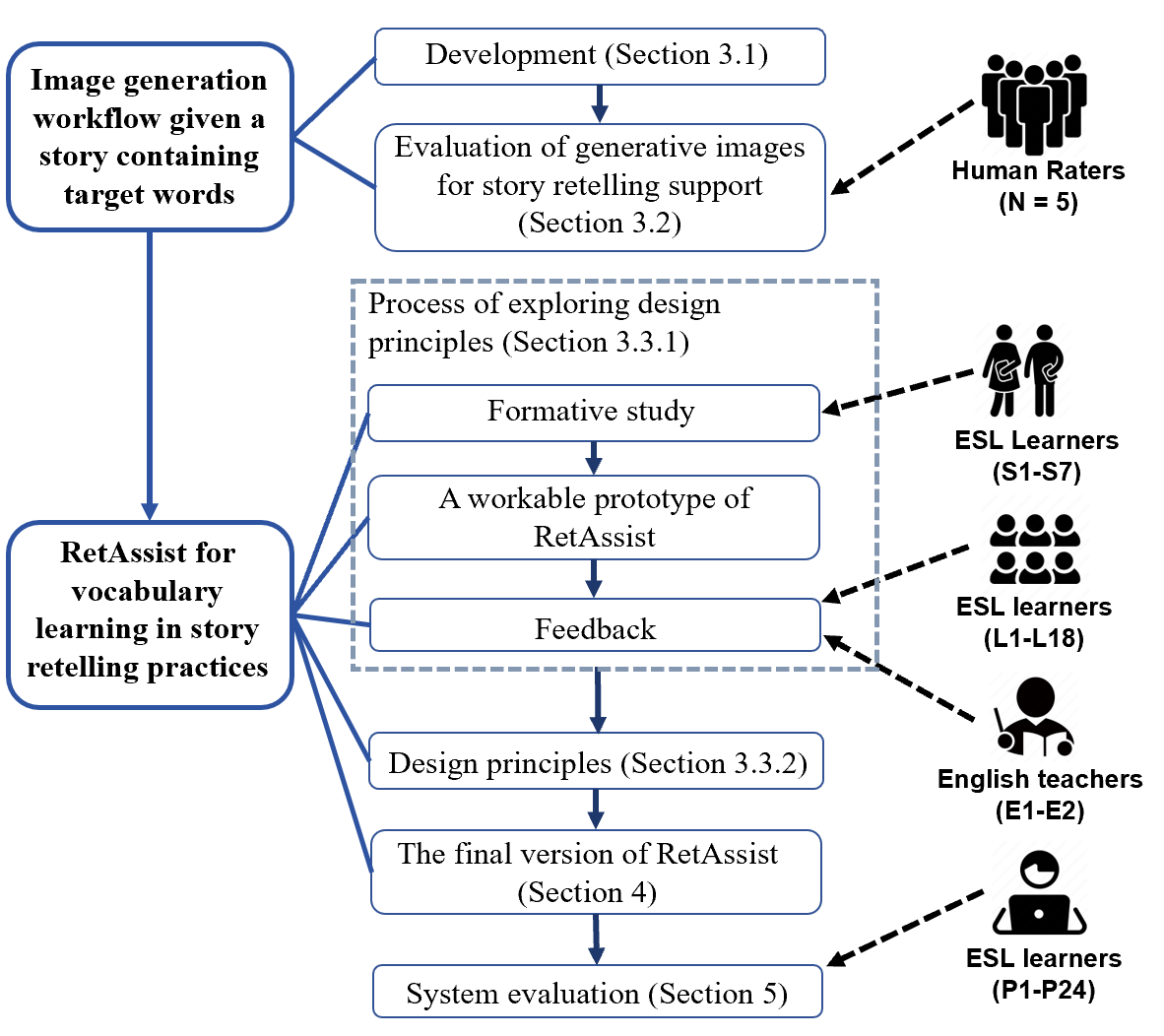}
    \Description[Flowchart]{Flowchart depicting the design and development process of \name{}, outlining the process from Section 3 to Section 5.}
    \caption{
    Our design and development process of \name{} with English teachers and ESL learners.
    }\label{process}
\end{figure}

\chen{
In this section, we explain how we design and develop \name{} to facilitate vocabulary learners to read and repeatedly retell any story that covers their interested target words (\autoref{process}). 
First, we propose a computational workflow for text-to-image generation and validate its feasibility in generating a series of coherent and relevant sentence-level images given any short textual story that contains target words. 
\zhenhui{Then, we work with vocabulary learners and teachers to derive the design principles of \name{}.}
% Then, we explore and come up with the design principles of \name{}. 
}

% As we anticipate \name{} can support retelling practices on any word set and associated story, we want to offer relevant images for any story. 
% Due to the lack of existing images that could satisfy our purpose, we develop a computational workflow for generating story-relevant image sequences.
% we choose to leverage the state-of-the-art text-to-image generation models. 
% Below, we first describe our applied techniques and then present the evaluation that validates the proper quality of the generative images for learning support.

\begin{figure*}[bp]
    \centering
    \includegraphics[width=14cm]{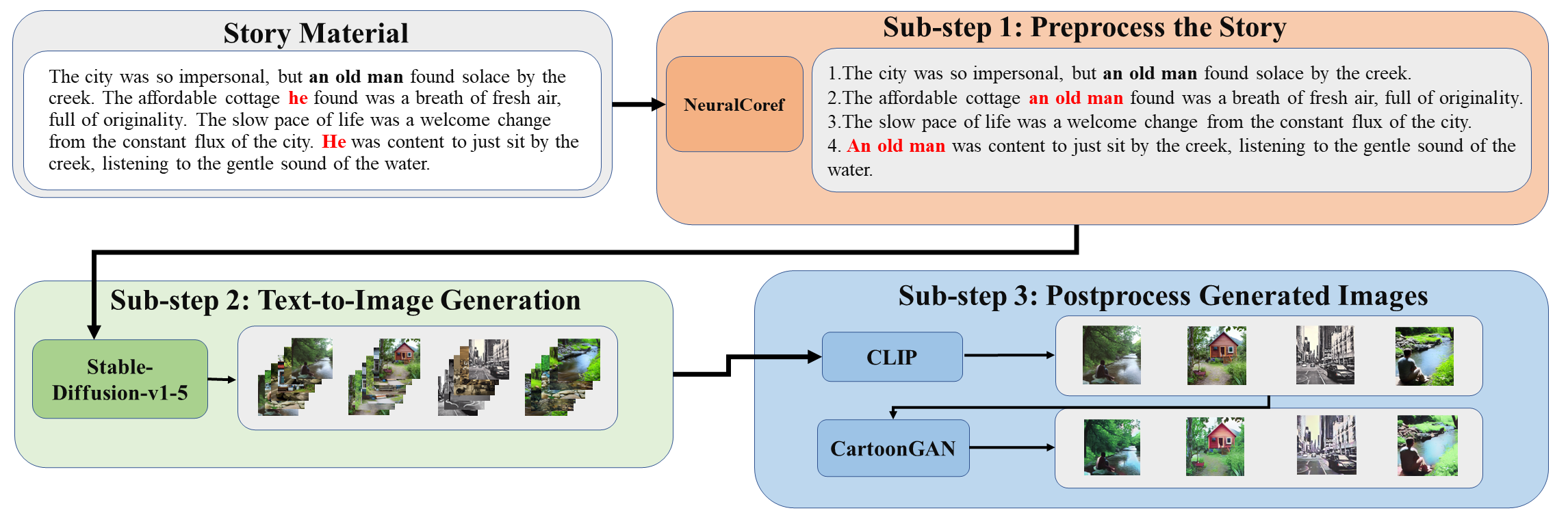}
    \Description[Flowchart]{Flowchart showing three steps in generating images from a story material, including preprocessing the story, text-to-image generation and postprocess generated images. }
    \vspace{-0.3cm}
    \caption{
    Our computational workflow of generating relevant images for stories.
    }\label{computational_workflow}
\end{figure*}

\begin{figure*}[htbp]
    \centering
    \includegraphics[width=14cm]{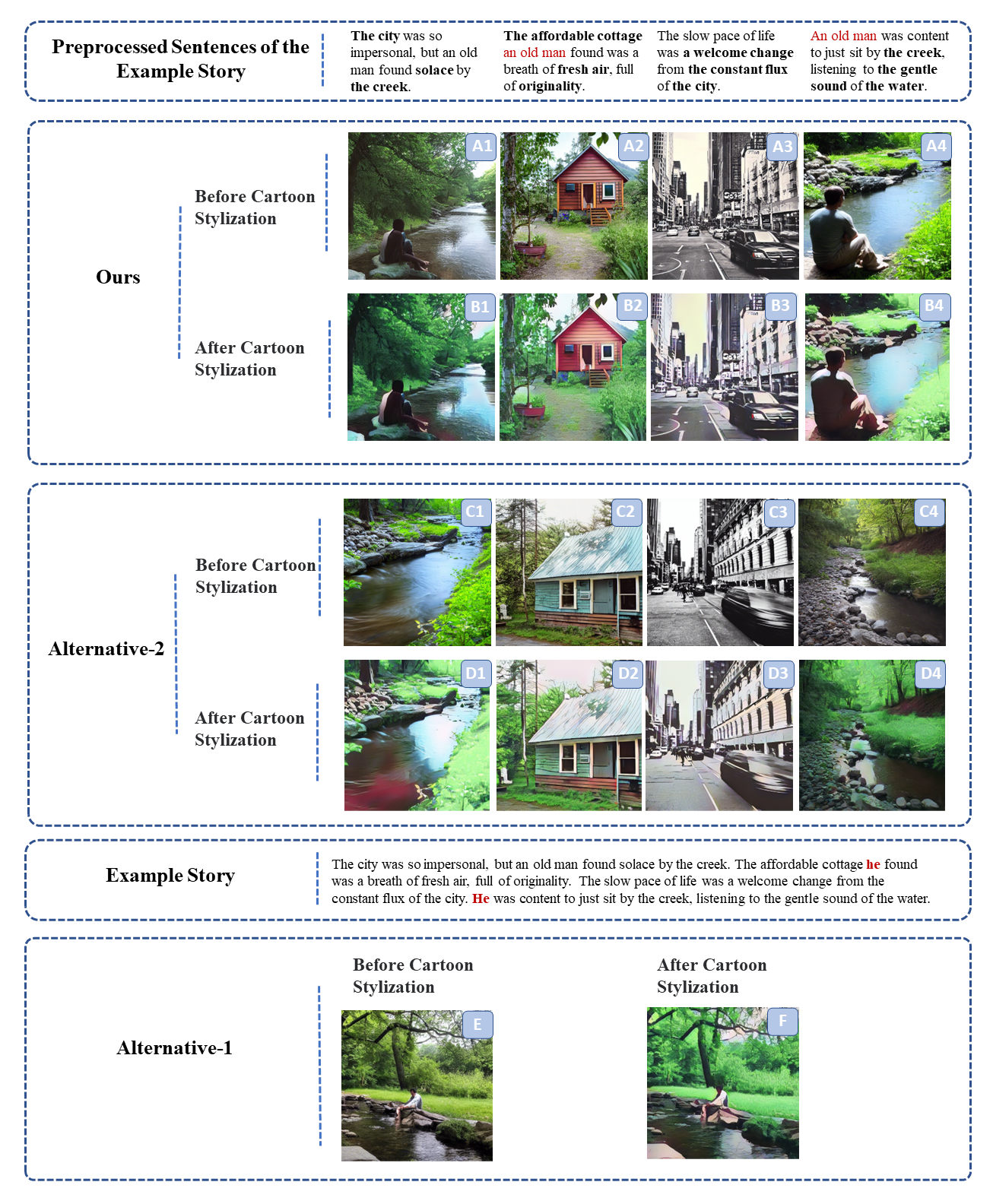}
    \Description[Visual comparison chart]{Visual comparison chart depicting images generated by Ours, Alternative-2 and Alternative-1 and corresponding cartoon stylization results. }
    \caption{
    % Comparsion of the story-relevant images generated by our approach with those generated by the alternative.
    Given sentences of an example story as input, we compare images generated by our computational workflow with those generated by two alternatives.
    [\textbf{Ours} (sentence-level, sentence-based)] A1-A4: Images generated using the preprocessed sentences as prompts. 
    B1-B4: Cartoon stylization of A1-A4. 
    [\textbf{Alternative-2} (sentence-level, keyword-based)] C1-C4: Images generated using the keywords (bold words in the preprocessed sentences of the example story) corresponding to the preprocessed sentences as prompts.
    D1-D4: Cartoon stylization of C1-C4. 
    [\textbf{Alternative-1} (story-level)] E: Images generated using the entire story as a prompt.
    F: Cartoon stylization of E. 
    }\label{visualization_example}
\end{figure*}

\peng{
\subsection{Developing a Computational Workflow for Text-to-Image Generation}
\label{subsection:genimg}
% Given the benefits of images for facilitating vocabulary learners in story retelling practices \cite{paivio1980dual, oktarina2020effectiveness, filippatou1996pictures, mayer2002multimedia, sorden2012cognitive, paivio1990mental, paivio2014bilingual}, we propose a computational workflow to generate relevant images given any short textual story. 
\cqyrevise{
\zhenhui{
To assist users in the story retelling practices, the generative images of a story should satisfy the following two requirements. 
First, the images should be semantically relevant to the textual story.}
As suggested by \qiaoyi{the Dual Coding Theory} \cite{paivio1980dual, paivio1990mental}, the brain processes visual and verbal information in distinct regions. The visual channel handles visual data, generating pictorial representations, while the verbal channel processes verbal information, producing corresponding verbal representations. 
% Based on consistent visual and verbal inputs therein, 
\zhenhui{When the visual and verbal inputs are semantically relevant, people} 
establish mental connections that organize information into cause-and-effect chains \cite{hasnine2021wordhyve}. After these connections are formed, there is a significant enhancement in the ability to remember information \cite{Peker2018TeachingAL}. 
Therefore, when serving language learning, the visual information in the image should semantically match with the text content.
\zhenhui{Second, the images themselves should be coherent in their content and consistent in styles to depict a story \cite{mayer2002multimedia}. Otherwise, the images could confuse learners in the story retelling practices.}
}
% Furthermore, the images should be consistent in their styles and reveal the coherent logic of the story \cite{mayer2002multimedia}. 
% As suggested by the Bilingual Dual Coding Theory \cite{paivio1980dual}, when serving for language learning, the visual information in the image should semantically match with the text content. 
% Furthermore, the images should be consistent in their styles and reveal the coherent logic of the story \cite{mayer2002multimedia}. 
% The goals of the workflow are, therefore, to generate relevant, content-coherent, and style-consistent images for the textual story. 
These two requirements guide our design choices in the computational workflow, as detailed below (\autoref{computational_workflow}). 
}
% \zhenhui{
% % [Peng: consider rewriting this paragraph like this:] 
% To assist users in the story retelling practices, the generative images of a story should satisfy the following requirements. 
% First, the images should be semantically relevant to the textual story. %, as the Bilingual Dual Coding Theory \cite{paivio1980dual, paivio1990mental} suggests that the consistent visual and verbal inputs to human's brain helps to 
% }
% shows our computational workflow consisting of three sub-steps. 
% There are three sub-steps to conduct image sequence generation for story materials. 
% First, we preprocess the story to obtain suitable sentences as input to the model to generate images. 
% Then, we build a text-to-image generation model that outputs multiple images for each input sentence. 
% Lastly, we postprocess the generative images to select and polish the associated images for the story. 
% As shown in Figure \ref{tech}, we first preprocess story materials to obtain story sentences that can be processed independently. Then, we use each full sentence as the text prompt of text-to-image generation. Finally, we postprocess the generative images for higher-quality image sequences.

\peng{
\textit{\textbf{Sub-step 1: Preprocess the Story.}}
% Aiming to get story sentences that can be processed independently, we preprocess story material by coreference resolution. 
% \zh{We choose to prepare one image for one full sentence as it could represent a complete story plot (? [what's your arguments for this design choice]).}
We choose to generate one image for each story sentence for two reasons. First, a series of images rather than one image can better reveal the logic of the story \cite{paivio1980dual}. %foster dual encoding of visual information with L1-L2 linguistic information \cite{paivio1980dual}. 
% Moreover, in the L2 learning approach based on CTML (Cognitive Theory of Multimedia Learning), supporting dual encoding of visual information with L1-L2 linguistic information during the L2 vocabulary learning phase may enhance recall of L2 vocabulary later on, as learners are able to remember L2 vocabulary based not only on L1 linguistic information, but also on shared visual information \cite{paivio1980dual}. 
Second, the Segmenting Principal \cite{mayer2002multimedia} suggests that preparing an image for each story segment can provide natural pauses for learners to absorb the content before proceeding to the next segment \cite{kalyuga1999managing}. 
We use the Spacy package in Python to split the story into sentences. 
% To generate a sequence of relevant images for a story that contains multiple sentences, we need to 
To maintain the coherence among the generative images, we further resolve coreferences
% address the coreference resolution problem 
in the story sentences, \eg pronouns like ``he'' and ``it'' refer to the objects mentioned earlier. 
% Coreference Resolution
% Pronouns are often used to refer to specific objects mentioned earlier in the later text, which is often the case with the story materials used in our study.
Specifically, we adopt a pretrained coreference resolution model named NeuralCoref \cite{wolf2018neuralcoref} to select the reference words in the story to replace the pronoun in each split sentence. 
For example, for the red text in  \autoref{visualization_example}, the pronounce ``he'' in the second and fourth sentences of the example story is replaced by ``an old man''. 
}
% selected by NeuralCoref. 
% The output of this sub-step is therefore a sequence of story sentences with the coreference problem resolved. 
% NeuralCoref \cite{wolf2018neuralcoref} is a data-driven NLP approach to selecting the most likely reference-word in the text for each occurrence of a pronoun.
% In the second and fourth story sentences of the story material in Figure \ref{tech}, the word "he" is used to refer to "an old man" having appeared in the first sentence, thus "he" is replaced by "an old man" after preprocessing.
% The result of this sub-step is shown in Figure \ref{replace}. 
% In order to ensure the integrity of the generative image for each sentence in the story, we use the pronoun coreference resolution method implemented by NeuralCoref \cite{wolf2018neuralcoref}.
% This data-driven NLP approach selects the most likely reference-word in the text for each occurrence of a pronoun.
% The result of this sub-step is shown in Figure \ref{replace}. 

%\textit{\textbf{Sub-step 2: Get Text Prompts for Image Generation}}
% The alternative is to directly use each sentence of the story as text prompts of image generation.
% }
% \pzh{
\xingbo{\textit{\textbf{Sub-step 2: Text-to-Image Generation.}}}
% Given the sentences from last step as input, we ... \zh{I am confused here, one sentence --> model --> multiple images? Or Multiple sentences --> model --> multiple images for each sentence? Or others? Please clarify.}
% }
% Given the sentences from the previous step as input, 
After preprocessing the story, we proceed to generate multiple images for each story sentence. 
% Having gotten story sentences that can be processed independently, the output of this sub-step is to generate multiple images to represent the sentence description.
% To obtain images that cover the main visual information of the story content, we utilize story sentences as prompts for image generation. 
% According to Segmenting Principal \cite{mayer2002multimedia}, users learn better when a multimedia lesson is presented in user-paced segments rather than as a continuous unit. Besides, Kalyuga \etal \cite{kalyuga1999managing} explore the effects of cognitive load and instructional design factors, such as pacing and segmenting information, on multimedia learning. 
% In the case of a short story with a limited number of sentences, segmenting it into individual story sentences can provide natural pauses for learners to absorb the content before proceeding to the next segment \cite{kalyuga1999managing}.
% Thus, we consider generating individual images in our computational workflow in units of one story sentence, \ie utilize the preprocessed sentences as prompts for image generation. 
Specifically, we leverage a state-of-the-art pretrained text-to-image generation model named Stable-Diffusion-v1-5, released by RunwayML and available in the Hugging Face model hub \cite{runwayml:stable-diffusion-v1-5}, \chen{because of its demonstrated capability to generate high-quality images relevant to the text \cite{Rombach_2022_CVPR}. }
%high inference spend and 
The model outputs five images given a preprocessed input story sentence. 
% To obtain images covering the main visual information as completely as possible, we suppose the entire story sentences are text prompts of text-to-image generation.
% To generate images from text prompts, 
% We use a certain trained model 

% \pzh{
\xingbo{\textit{\textbf{Sub-step 3: Postprocess generative images.}}}
With the candidate generative images, we further select and polish the most relevant image for each story sentence.
% Given the multiple generative images for each sentence in the last step as input, this step selects the most relevant image for each sentence and polishes the selected images. 
% In this sub-step, the images generated in the previous sub-step are postprocessed to produce an image sequence representing the story material.
% \zh{xxx [short description of this mode]} 
The selection is based on the semantic similarity between the sentence and its candidate images. 
Specifically, we use a pretrained cross-modal model named CLIP \cite{radford2021learning} to encode the sentence and image into vectors and compute the cosine similarity of the image.
% To measure the relevance between generative images and story sentences, we utilize a pretrained cross-modal model, CLIP \cite{radford2021learning}.
After selecting the images (\eg A1-A4 in \autoref{visualization_example}) with the highest similarity scores with the story sentences, we seek to mitigate the potential inconsistencies among the selected images, 
% \eg the same human character can have varied visual appearances across images.
\qiaoyi{\eg the same human character may be visually represented differently across images, such as variations in hair or facial details. }
% It encodes images and texts into high-dimensional vectors, which depict the relationships between them. 
% We use a trained model CLIP (Contrastive Language-Image Pre-Training), which associates images and their descriptions, to encode each story sentence and corresponding generative images into vectors \cite{radford2021learning}.
% As CLIP (Contrastive Language-Image Pre-Training) is a neural network trained on a variety of (image, text) pairs, we used it to encode sentence and corresponding generative images for each story sentence.
% Then, for each sentence, we calculate the cosine similarities between its vector and image vectors and select the image with the highest similarity score. %and then select the most relevant image for each story sentence.
% Although we have reduced the importance of consistency in determining which images correspond to the story sentences,
% For example, the images E1-E4 in Figure \ref{visualization_example} are the selected images for each sentence of the example story using our approach. 
% \zh{Confusing, B1-B4 are also generated by entire story sentences} 
% Moreover, there may exist inconsistencies between images (\eg same human character can have varied visual appearances across images). 
We adopt a cartoon-style transfer model \cite{chen2018cartoongan} that can convert each image to match a cartoon style while maintaining the original structures, textures, and basic colors of the image (\eg B1-B4 in \autoref{visualization_example}). 

\subsection{Evaluating the Feasibility of Generative Images for Story Retelling Support}
At this stage, we would like to compare the quality of the images generated by our workflow with those generated by alternative approaches given the same short story. 
This evaluation aims at validating if the generative images are relevant to the story, have acceptable visual quality, and are perceived as helpful in helping learners comprehend and recall the story. % competent for supporting learners in the story retelling practices. 
We will assess the effectiveness and user experience of generative images in story retelling in the later experiments with vocabulary learners. 
Inspired by prior work on text-to-image generation \cite{koh2021text}, story-related images generation \cite{li2019storygan}, and the usage of images in story retelling practices \cite{dunst2012children}, we derive the following evaluation metrics: 
\textbf{relevance} (\textit{The images are relevant to the story description}), \textbf{visual quality} (\textit{The images are close to the real scene}), \textbf{perceived effectiveness in aiding comprehension} (\textit{The images are helpful if you are going to do story comprehension}), and \textbf{perceived effectiveness in aiding recall} (\textit{The images are helpful if you are going to do repeated retelling}).
Each item is rated on a standard five-point Likert Scale (1 for ``Strongly disagree'' and 5 for ``Strongly agree'').

\peng{
\subsubsection{Alternative approaches}

We compare our computational workflow with two alternative approaches for text-to-image generation. 
The first one, noted as \textbf{Alternative-1}, generates ten images by directly inputting the original story to the Stable-Diffusion-v1-5 model and then selects and stylizes the most relevant one (\eg F in \autoref{visualization_example}) similar to the sub-step 3 in our workflow. %based on whole story without preprocessing 
\qiaoyi{
Alternative-1 produces a single image for the entire story refer to \textit{CoSpeak} \cite{10.1145/3462204.3481750}, which provides a single image to assist two English learners to co-create a story through dialogue.
}
The comparison with Alternative-1 (\ie sentence-level vs. story-level) aims at checking if generating a series of sentence-level images could be more helpful than generating one story-level image. 
The second approach, noted as \textbf{Alternative-2}, uses TextRank to extract keywords (\eg the bold ones in \autoref{visualization_example}) as prompts to the Stable-Diffusion-v1-5 model to generate five images \cite{liu2022design}. 
It then selects and polishes the most relevant image (D1-5 in \autoref{visualization_example}) for each sentence using the same postprocess methods in our proposed workflow. 
By comparing our workflow to Alternative-2 (\ie sentence-based vs. keyword-based), we aim to examine if the sentence-based prompt would be better than the keyword-based prompt, \chen{as a related work suggests that these two prompts were comparable in text-to-image generation tasks \cite{liu2022design}. }
% As for Alternative-1, we use the entire original story as a single prompt for generating images corresponding to the whole story. 
% We then postprocess the generative images with the same as Our proposed computational workflow. 
% The images A in Figure \ref{visualization_example} show an example of how the single image generated by Alternative-1 look like. 

% As for Alternative-2, we use TextRank to extract keywords for each sentence (The keywords in the example story are bolded in the preprocessed sentences in Figure \ref{visualization_example}) as prompts to generate images \cite{liu2022design}, instead of directly using the preprocessed story sentences as prompts for generation. 
% We compare the keyword-based alternative technique with the sentence-based approach since a related work suggests that they were comparable in text-to-image generation tasks [A1]. 
% Then, we postprocess the generative images with the same as Our proposed computational workflow. 
% The images D1-D4 in Figure \ref{visualization_example} show an example of how the image sequence generated by Alternative-2 look like. 
}

% Apart from the technical evaluation, we further conduct two human evaluations to validate the feasibility of generative images for supporting story retelling practices. 
% Respectively, we compare our proposed computational workflow (denoted as \textbf{Ours}) with the two aforesaid alternatives (denoted as \textbf{Alternative-1} and \textbf{Alternative-2}).
% We propose four evaluation metrics based on previous work on text-to-image generation \cite{koh2021text} and story-related images generation \cite{li2019storygan}, as well as our application scenario that uses images to assist story retelling practices \cite{dunst2012children}. 
% The metrics include \textbf{relevance} (\textit{The images are relevant to the story description}), \textbf{visual quality} (\textit{The images are close to the real scene}), \textbf{effectiveness in aiding comprehension} (\textit{The images are helpful if you are going to do story comprehension}), and \textbf{effectiveness in aiding recall} (\textit{The images are helpful if you are going to do repeated retelling}).
% Each item is rated on a standard five-point Likert Scale (1 for ``Strongly disagree'' and 5 for ``Strongly agree''). 

\peng{
\subsubsection{Preparing target word sets and short stories}
\label{subsubsec:stories}
We prepare 20 short stories, each containing a given target word set, to compare the images generated by our proposed workflow with those generated by alternative approaches. 
% \textbf{Target Word Sets. } 
% Since the target users of \name{} are university students in China, We anticipate that \name{} can support story retelling practices on any target word IELTS set that learners or teachers specify. 
The target words are from the vocabulary pool (3,672 words in total) suggested by the International English Language Testing System (IELTS) \cite{cullen2012cambridge}. 
Three authors of this paper randomly select non-easy IELTS words (\eg not the words like ``easy'' and ``general'') that they did not know before, which are randomly assigned to 20 sets, each with six or seven words. %(have passed CET-6 \footnote{https://cet.neea.edu.cn/ The College English Test-6 (CET-6) is a standardized English proficiency exam in China, administered by the Ministry of Education.})
% ----------------------------- %
% Each vocabulary lists including 6-7 keywords from the vocabulary pool (3,672 words in total) suggested by the International English Language Testing System (IELTS) \cite{cullen2012cambridge}. 
% Specifically, these keywords are difficult vocabulary that screened out by three authors (have passed CET-6 \footnote{https://cet.neea.edu.cn/ The College English Test-6 (CET-6) is a standardized English proficiency exam in China, administered by the Ministry of Education.}) . 
% \textbf{Short Stories. }
This manipulation simulates the case in which learners would like to learn any interested target word set via story retelling. 
To prepare a story for each target word set, we first query ChatGPT \cite{bang2023multitask} with ``generate a short story that has no more than 60 words and must contain the words `[word 1]', `[word 2]', ..., and `[word n]' ''.  %''use the words `[word 1]', `' to generate a simple story and limit the story to 60 words.''  %to generate 20 concise stories featuring words from vocabulary list by requesting 
This approach leverages the capability of the recent large language models to generate a short story that contains any target word set \cite{peng2023storyfier}.  
% \qiaoyi{Compared to using existing short stories validated by English teachers, stories generated by ChatGPT can be customized to suit learners' needs and interests on mastering any target words.  %, offering a more personalized approach. 
% }
\chenqy{
Compared to using existing short stories validated by English teachers, stories generated by ChatGPT can be flexibly adapted to learners' needs and interests on mastering any target words.
}
% We anticipate that \name{} can support story retelling practices on any target word IELTS set that learners or teachers specify, so we utilize the state-of-the-art technology ChatGPT instead of searching for existing stories as retelling materials.
% Furthermore, we simplify the generated story to improve the readability using ''rewrite the story with subject-predicate structure'' if necessary. 
The first author then refines the generated stories to improve their readability. 
% For instance, ``The heavy rain exacerbated the flooding in the low-lying areas'' is better understood than ``The flooding in the low-lying areas was exacerbated by the heavy rain''. 
% After this step, 
Finally, we get 20 short stories (average word length: 60, average number of sentences: 5) that cover topics like funny animals, disasters, everyday life, and travel. 
% ----------------------------- %
}

\peng{
\subsubsection{Procedure and Results}
% For the first human evaluation, we would like to explore whether a single image generated based on the entire story is more suitable than multiple images generated corresponding to preprocessed story sentences for our story retelling practices.
\
\newline
\textbf{Vs. Alternative-1}. We prepare a document that lists the 20 stories; following each, there is a series of images generated by our workflow, an image generated by Alternative-1, and spaces for raters to input their scores for each metric.
% We prepare a word document that lists the 20 stories, after each there are a series of images generated by our workflow, an image generated by Alternative-1, and the blanks for raters to input their scores on each metric. 
% Corresponding to each story, there are a single image generated by \textbf{Alternative-1} (\eg images A in Figure \ref{visualization_example}) and images generated by \textbf{Ours} (\eg images B1-B4 in Figure \ref{visualization_example}).
We distribute this document to five human raters (3 males, 2 females, age: $Mean = 20.6, SD = 0.49$) recruited from a local university. % to assess the 20 image sets in a shuffled order. 
% For each item in each sequence, we average the human raters' ratings as the final score, and we use paired samples wilcoxon signed rank tests to detect significance. 
For each metric of the generative image(s) for a story, we average the scores of five raters as the final score. 
Next, we use paired-sample Wilcoxon signed rank tests to analyze the differences between our workflow and Alternative-1 on each metric. %  \chen{(N = 5)} 
% Then, we generate 20 sets of images corresponding to the stories via both Strategy 1 and Strategy 2, and all of these image sequences of stories are assigned to 5 workers (3 males, 2 females, age: $Mean = 20.4, SD = 0.27$) to reduce human variance. 
% The order of the options within each assignment is shuffled to make a fair comparison. 
% All human raters are undergraduates in a local university who habitually read English texts and follow their illustrations. 
% The images are evaluated regarding \textbf{relevance} (\textit{The images and sentences of the story are very relevant}), \textbf{visual quality} (\textit{The set of images is in line with real scenarios}), \textbf{effectiveness in aiding comprehension}, and \textbf{effectiveness in aiding recall} on a standard five-point Likert Scale (1 for “Strongly disagree” and 5 for “Strongly agree”). 
% Each rater is compensated with 2.92 USD. 
% results
% As shown in ...
% The results of the human evaluation study are summarized in Figure i. 
% The mean values of the ratings for all 20 stories indicate that Strategy 2 outperforms Strategy 1 in terms of quality overall on the task of story visualization with \name{}.
% Table \ref{human_tab1} shows the average scores of two computational techniques for generating images associated with the stories in our rated 20 sets of images.
\qiaoyi{
As depicted in \autoref{image_rate1}, our workflow performs significantly better in generating relevant image(s) to the story than the Alternative-1 ($p < 0.05, z = 2.023, \text{Cohen's d} = 1.208$). 
Raters also perceive that our generative images are significantly more effective in aiding comprehension ($p < 0.05, z = 2.023, \text{Cohen's d} = 1.417$) and recall of the story ($p < 0.05, z = 2.023, \text{Cohen's d} = 1.537$). % than \textbf{Alternative-1}, and no significant difference in visual quality. 
}
These results indicate that generating a series of sentence-level images about a story could be more helpful in story retelling than generating one story-level image.
% Thus, multiple images generated corresponding to preprocessed story sentences are generally superior to a single image generated based on the whole story. 

\begin{figure*}[htbp]
    \centering
    \begin{minipage}[t]{0.48\textwidth}
        \centering
        \includegraphics[width=7cm]{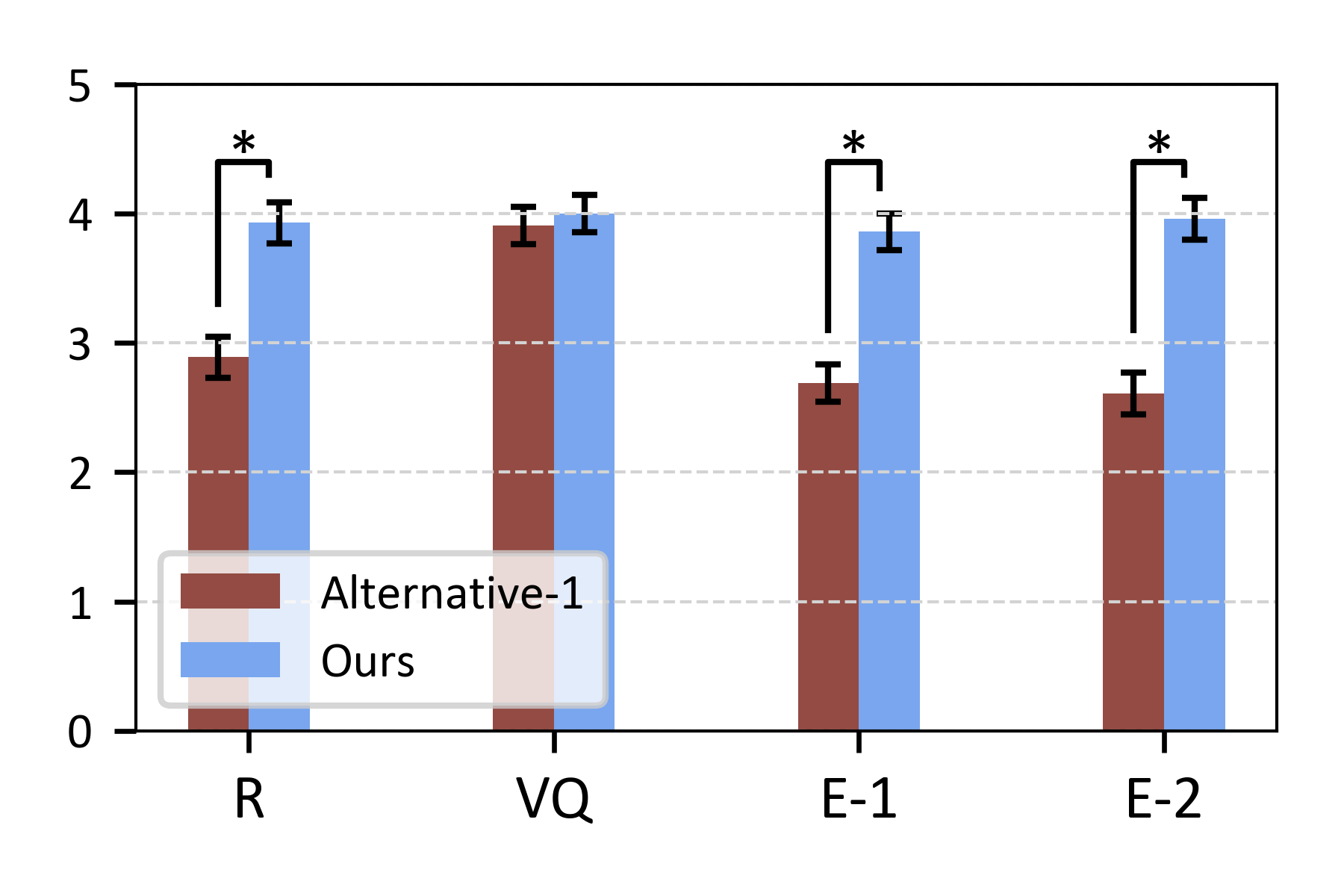}
        \Description[Bar graph]{Bar graph comparing human ratings of generative images and there exists significant difference on R, E-1 and E-2. }
        \vspace{-0.5cm}
        \caption{
    Means and Standard Errors of human ratings on the quality of generative images; 1/5 - strongly disagree/agree; \qiaoyi{*: $p$ < .05 using paired samples Wilcoxon signed rank tests.} We compare \textbf{Alternative-1} (story-level) with \textbf{Ours} (sentence-level) on the images’ relevance (R) to the story, visual quality (VQ), and effectiveness in aiding story comprehension (E-1) and recall (E-2). 
    }\label{image_rate1}
    \end{minipage}%
    \hspace{0.2cm}
    \begin{minipage}[t]{0.48\textwidth}
        \centering
        \includegraphics[width=7cm]{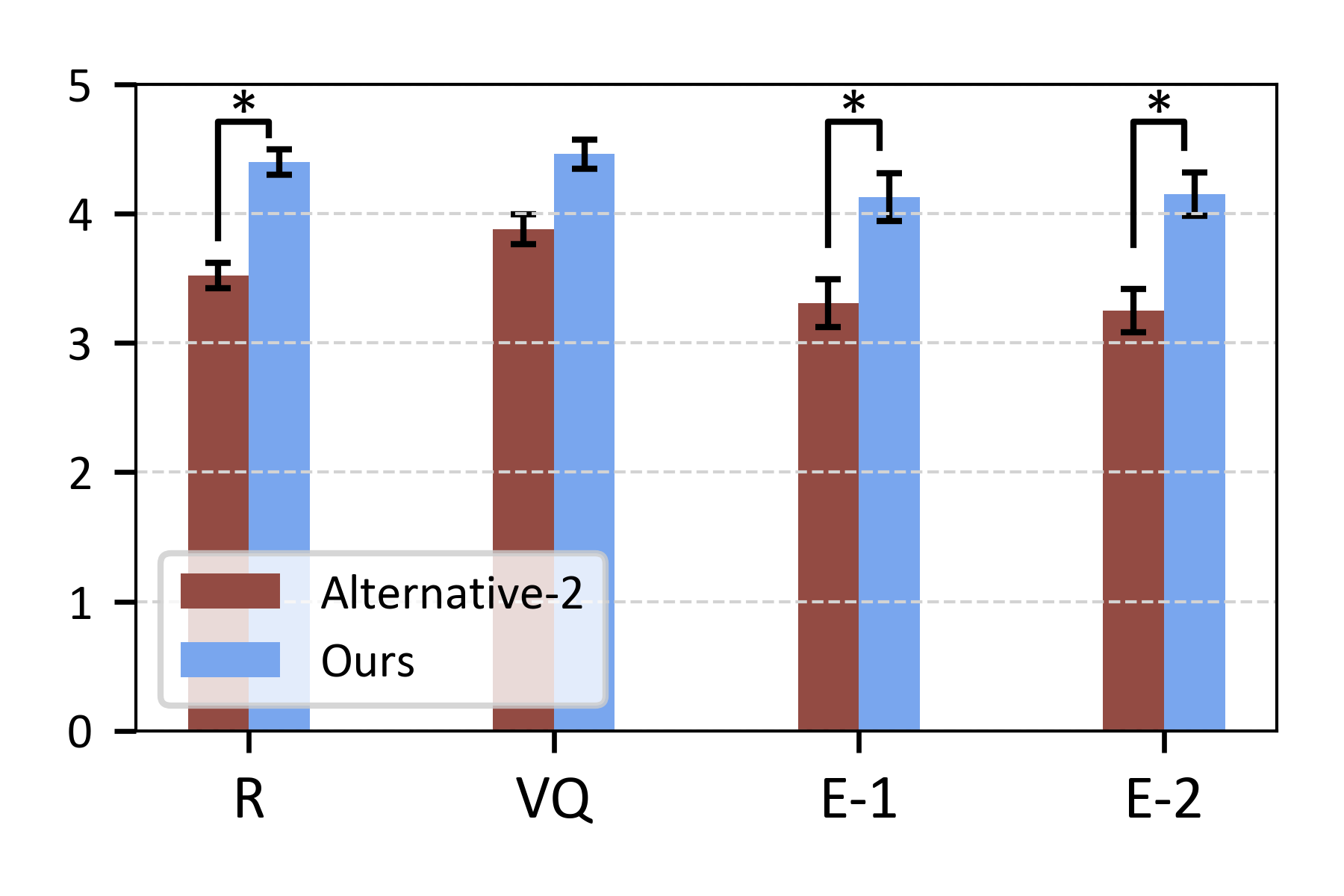}
        \Description[Bar graph]{Bar graph comparing human ratings of generative images and there exists significant difference on R, E-1 and E-2.}
        \vspace{-0.5cm}
        \caption{
    Means and Standard Errors of human ratings on the quality of generative images; 1/5 - strongly disagree/agree; \qiaoyi{*: $p$ < .05 using paired samples Wilcoxon signed rank tests.} We compare \textbf{Alternative-2} (keyword-based) with \textbf{Ours} (sentence-based) on the images’ relevance (R) to the story, visual quality (VQ), and effectiveness in aiding story comprehension (E-1) and recall (E-2). 
    }\label{image_rate2}
    \end{minipage}%
\end{figure*}

\textbf{Vs. Alternative-2}. 
% For the second human evaluation, we would like to explore whether extracting keywords from sentences as prompts \cite{liu2022design} for image generation is more practical than naively using the preprocessed sentences as prompts. 
% We compare the keyword-based alternative technique with the sentence-based approach since a related work suggests that they were comparable in text-to-image generation tasks [A1]. 
% Corresponding to each story, there are images generated by \textbf{Alternative-2} (\eg images D1-D4 in Figure \ref{visualization_example}) and images generated by \textbf{Ours} (\eg images B1-B4 in Figure \ref{visualization_example}). 
Similar to the procedure in comparing with Alternative-1, we recruit another five human raters (3 males, 2 females, age: $Mean = 20.4, SD = 0.27$) from the local university to score the images generated by our workflow and Alternative-2 on each metric \qiaoyi{and conduct paired-sample Wilcoxon signed rank tests. }
The order of encountering images of each story in the rating document is randomized and blind to the raters. 
%assess the 20 image sets in a shuffled order, and we use the same approach as the former human evaluation to processing the ratings and detecting significance. 
% For each item in each sequence, we average the human raters' ratings as the final score, and we use paired samples wilcoxon signed rank tests to detect significance. 
\qiaoyi{
As shown in \autoref{image_rate2}, compared with the Alternative-2, images generated by our workflow are significantly more relevant to the story ($p < 0.05, z = 2.023, \text{Cohen's d} = 1.809$) and are perceived significantly more effective in aiding comprehension ($p < 0.05, z = 2.032, \text{Cohen's d} = 0.872$) and recall ($p < 0.05, z = 2.032, \text{Cohen's d} = 1.06$) of the story.
}
% ($p < 0.05$, paired-sample Wilcoxon signed rank tests). %, and is marginally better in visual quality than \textbf{Alternative-2}. 
% Therefore, images generated corresponding to preprocessed story sentences are overall better than images generated corresponding to keywords extracted from each sentences. 
These results indicate that generating a series of sentence-level images about a story using the sentence-based prompt could be more helpful in story retelling than generating the image series using the keyword-based prompt.

To sum up, the results of the evaluation study support our choices to generate sentence-level images using sentence-based prompts. % the results of two human evaluations, images generated by \textbf{Ours} generally outperform the two alternatives. 
% Besides, the means of four evaluation metrics in our computational workflow are all larger than or equal to 4.00 out of 5 points. 
The means of the four metrics on the images generated by our computational workflow are all larger than or equal to 4 out of 5 points, indicating its feasibility for generating images that are relevant to the story, of high visual quality, and potentially helpful to support story retelling. 
We then proceed to explore how our generative images can be used to support vocabulary learners in their story retelling practices. 
% We will evaluate the effectiveness of these images for vocabulary learning in the later user study.
% This means that the generative images are highly relevant to the stories and the text-to-image generation techniques of \textbf{Ours} are feasible in facilitating vocabulary learners in their story retelling practices. 
}
% indicating the feasibility of text-to-image generation techniques for vocabulary learning via story retelling. 
% Feasibility and Support of Text-to-Image Generation Techniques for Vocabulary Learning via Story Retelling. 

\cqyrevise{
\subsection{Exploring Design Principles of \name{}}
% \subsection{Exploring Design Requirements of Using Generative Images in Story Retelling Practices}
% \subsection{Exploring Story Retelling Support with Generative Images}
% After validating the feasibility of our computational workflow for generating images that are relevant to the story and have acceptable visual quality, we explore possible activities that the generative images can support in story retelling practices. 
\zhenhui{With our computational workflow for text-to-image generation as the backbone of \name{}, we work with vocabulary learners and teachers to derive design principles of \name{}.}
% In this subsection, we will explain how we come up with the design principles of \name{} that use generative images to support vocabulary learners in story retelling practices.

\subsubsection{Process of exploring design principles}
To put forward design principles on how to build a vocabulary learning system that uses generative images in story retelling practices, we first conduct a formative study with seven ESL (English-as-Second-Language) learners.
\zhenhui{Then, we develop a workable prototype of \name{}.}
% then develop a workable prototype of \name{} based on the results of the formative study. 
Next, we evaluate the \name{} prototype through a within-subjects study with 18 ESL learners. According to user feedback on the prototype, we prepare a revision plan on \name{} and solicit feedback from two English teachers. 
% \zhenhui{The details of this process are as follows.}
% The stages of the design principles exploration process are specifically described as follows.

\textbf{\textit{Formative study with seven ESL learners.}}
% [One sentence about the goal and participants -- To receive an initial understanding of the needs and requirements of learners for a ... system, we ... (basic participant information)]
% To receive an initial understanding of the needs and requirements of ESL learners for a vocabulary learning system based on story retelling practice, 
\zhenhui{To understand user needs and requirements for a system that provides generative images in the story retelling practices,}
we conduct a formative study with seven ESL college students (S1-S7, 1 male, 6 females, age: $Mean = 20.57, SD = 0.82$) in China. 
% [1-2 sentences about content of the study -- two story retelling practices, one with ..., one without ..., followed by questions about ...].
\cqyrevise{
Focusing on gathering the feedback and suggestions of ESL learners, we do not specifically balance the order of retelling with and without generative images in this instance. 
}
We first invite them to conduct one story retelling practice with generative images \footnote{The stories and images are listed in a Word file and come from the materials used in the evaluation of our workflow in \autoref{subsubsec:stories}.} and the other without images. 
Then, we ask questions about their perceptions of the practices and their expectations for a system using generative images to support story retelling. 
\qiaoyi{The findings here underpin DP1, DP2, DP4 and DP5 in \autoref{sec:design_principles}. }
% We invite them to conduct story retelling practices with both with-image system and without-image system, followed by questions about their perceptions of the story retelling practices and their expectations for an intelligence system that supports vocabulary learning with story retelling. 
% [1-2 sentences saying that we identify that the system can use generative images to facilitate comprehension of the story and help recall it when user gets stuck at each round of repeated retelling]
% [system can use generative images to facilitate comprehension and recall]
% These ESL learners favor the practice with generative images related to the story, considering that these images can help them better understand the story and recall the story line. %, and organize their . 
% [cloze prompts when get stuck]
% They expect that the intended system could provide prompts about the story when they get stuck in the repeated retelling stage. 
% [feedback]
% Besides, five of them want to get feedback on their performance in the practice, \eg about the usage and expression of target words. % so as to help reflect on each practice.  

\begin{figure}[htbp]
    \centering
    \includegraphics[width=11cm]{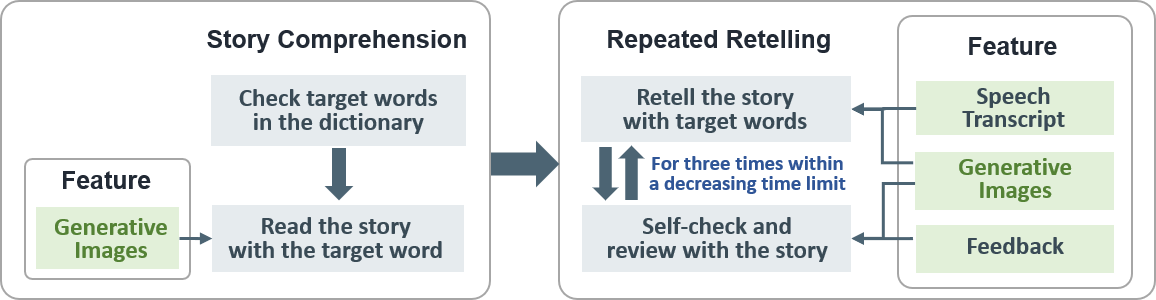}
    \Description[Flowchart]{Flowchart illustrating the structured story retelling practice flow, integrated with features like generative images, speech transcript, and feedback. }
    \caption{
    \pzh{The structured story retelling practice flow with the story comprehension and repeated retelling stages in \name{} and baseline systems. In the evaluation of \name{} prototype, the baseline system does not have features of speech transcript, generative images, and feedback. In the user study of the final version of \name{}, the baseline system does not have the generative images but has other features like \name{}.  }
    % The side parts are the proposed features of two systems, and the generative image is distinctive in \name{}. }
    % The system workflow. The upper part shows the flow of story retelling for vocabulary learning, including story comprehension and repeated retelling.
    }\label{fig:1}
\end{figure}

\textbf{\textit{Prototype of \name{}}.}
Based on the results of the formative study, we develop a workable prototype of \name{}. % to test different design hypotheses with end users to learn more about interface design and interaction with the generative images. 
This prototype structures the procedure of story retelling practice as used in the final \zhenhui{version of}  \name{} (\autoref{fig:1}, detailed in \autoref{sec:system_implementation}) but has several features different from the final \zhenhui{version of} \name{}.  % and \autoref{subsec:repeated_retelling}
For example, the generative images are sequentially fixed in the interface and are not interactive. 
Inspired by the study of Gu \etal \cite{gu1996vocabulary}, this prototype will prompt the next sentence that masks the keyword when users get stuck for five seconds during the repeated retelling stage. 
% [give one or two main different features and our goals to test them]. 
% 1) \name{} lists all generative images related to the story in order (\ie users don't need to have any interaction with the images), so as to validate whether showing all the images statically is an appropriate way to offer visual materials.  
% 2) In order to test whether the time-triggered prompts lead to users' dependencies, \name{} pops up a cloze sentence prompt based on users' retelling progress to help them proceed with their retelling practice. 
Besides, after each round of repeated retelling, this prototype provides feedback about the incorrect use of target words and the associated sentence but does not provide the story and generative images for review before the next round of retelling. 
\zhenhui{These features are discarded or refined in the final version of \name{} based on the feedback from ESL learners and English teachers, as discussed in \autoref{sec:design_principles}.}
% 3) Aiming to propose an effective form to provide feedback for users, \name{} offers adaptive feedback in an interface separate from the retelling interface, which only provides the meanings of sentimentally incorrectly used target words as well as their corresponding story sentences.  

\textbf{\textit{Evaluation of the \name{} prototype}.}
% [One sentence about the goal and participants -- To probe the user experience of \name{} and get user feedback to improve it, we ... within-subjects study with ... (basic participant information)]
To probe user experience of the \name{} prototype and feedback to improve it, we conduct a within-subjects study with 18 ESL learners (L1-L18, 14 females, 4 males, age: $Mean = 20.56, SD = 1.17$). 
The task and procedure are similar to the later user study of the final \name{} (detailed in \autoref{subsec:procedure}), except that we do not have the pretest and the two posttests in this study.
% [1-2 sentences saying that we measure the user experience and get qualitative feedback on ...] 
% In the evaluation, we measure users' learning outcomes and experience. 
During the within-subjects study, we get their qualitative feedback on how the features of the \name{} prototype affect their learning process. 
% [1-2 sentences about the key things we learn from the study, what is good, and what features are not due to ...]
We compare our \name{} prototype and a baseline system without generative images, speech transcription, adaptive prompts, and feedback to explore the necessity of these system features. 
\cqyrevise{
Consistent with the user study of final \name{}, we counterbalance the order of the used systems and encountered word sets using Latin Square.
}
After the learning sessions, we ask about their experience in the story retelling practices, their perception towards the two systems, and suggestions for improvement. 
\qiaoyi{The findings here underpin DP1 - DP5 in \autoref{sec:design_principles}. }
% Based on the results, the majority of learners agree with the effectiveness of the generative images and consider that adaptive feedback can not only help them focus on the unskillful part but also make them feel progress intuitively. 
% In general, all learners favor the \name{}'s features like generative images, speech transcription, and feedback. 
% Nevertheless, eight learners comment that the proactive sentence prompts during the repeated retelling stage often interrupt their retelling process and may result in their dependence on the prompts to finish the retelling. 
% It is because they felt that thinking fully during the retelling practice is more important than completing the retelling, and that the prompt may interrupt their thinking and make them more inclined to wait for the prompt next time, thus resulting in dependence.
% Seven learners suggest that \name{} had better present the feedback together with the story and images, so that they can better review their performance in the last round of repeated retelling before proceeding to the next round. % while identifying incorrect usage.
% Besides, five learners suggest that the images should align with the story content in a more clear way. 
% Moreover, some believe that showing an entire sequence of images at once can lead to confusion in recall (\ie too much textual information is momentarily associated with the images). 

\textbf{\textit{Feedback from two English teachers}.}
Based on the user feedback, we prepare a revision plan in a PowerPoint file that draws possible designs for features about the interaction with generative images, prompts in the retelling stage, and feedback on user performance. 
We bring this plan and our \name{} prototype to two English teachers (E1, female, age:27; E2, male, age: 27) and ask for their critiques and suggestions. 
\qiaoyi{The findings here underpin DP1 - DP5 in \autoref{sec:design_principles}. }

\subsubsection{Design principles}\label{sec:design_principles}
We finalize five design principles (\textbf{DP}s) based on \zhenhui{the results from the design process}. 
% Finally, we come up with five design principles (\textbf{DP}s) on how to build a vocabulary learning system that uses generative images in story retelling practices: 
% To design an interactive system to facilitate vocabulary learning via story retelling practice,  
% \textbf{A table that describe the design principles that exactly match to our final \name{}}
% 1. images in the stage of story comprehension
% 1) Provide generative images during the story comprehension stage to aid users in better understanding the storyline and constructing connections between text and images. 

\textbf{DP1: In the story comprehension stage, \zhenhui{the system should} provide generative images to facilitate users in understanding and remembering the storyline. }
Previous educational literature suggests that images depicting the story could help learners to understand stories efficiently \cite{oktarina2020effectiveness, filippatou1996pictures}.
Our participants in the formative study favor the condition with images in their story retelling practices since the generative images can help them quickly and correctly understand the story.
``The associated pictures with the story help me understand and remember the storyline, enabling more efficient story retelling practices'' (S7).
Also, all learners in the evaluation study of the \name{} prototype expressed their \zhenhui{favor} for the generative images.
``I like to incorporate images to understand the story'' (L8). %favoritism
Both English teachers believe that generative images are practical materials to \zhenhui{promote story comprehension.} % support textual comprehension.
% \textbf{DP1}) In the story comprehension stage, provide generative images to facilitate users in understanding and remembering the storyline.  
% [When clicking on a particular image, the corresponding sentence should be emphasized to establish a clear association]
% [When clicking on a specific image, there should be alignment with the corresponding sentence to establish a clear association.]
% 2. images in the stage of repeated retelling
% 2) Provide image prompts during the repeated retelling stage to aid users in recalling the storyline based on the connections.
% 2) Implement image prompts in the repeated retelling stage to assist users in recalling the storyline by leveraging established connections, thereby enhancing memory retention and learning outcomes.

\textbf{DP2: During each round of the repeated retelling stage, \zhenhui{the system should} offer the generative images to help users recall the storyline yet not prompt the next sentence when users get stuck. }
As suggested by the Cognitive Theory of Multimedia Learning \cite{paivio1980dual}, visual elements (\eg figures) associated with the story can facilitate recall of words and their contextualized usage \cite{oktarina2020effectiveness, filippatou1996pictures}.
The ESL learners participating in both the formative study and the \zhenhui{evaluation} study indicate that the images can assist them recall the story and organize their retelling flow. %
``I can easily connect the pictures back to the story plot when retelling the story'' (S2).
``I connect the \zhenhui{images} provided with the story, which helps me reflect on the storyline in a short time'' (L3). %RetAssist
Previous learning support system \textit{EnglishBot} \cite{ruan2021englishbot} offers Chinese prompts to users in their conversations with a chatbot, and our participants in the formative study raise similar expectations that the intended system could provide in-situ prompts about the story when they get stuck in the repeated retelling stage.
``Rather than re-reading the full story, I’d like to get hints from the system about what’s next when I get stuck in the retelling'' (S3).
Nevertheless, as indicated by eight participants in the \zhenhui{evaluation} study of \name{} prototype, the proactive sentence prompts during the repeated retelling stage often interrupt their retelling process and may result in their dependence on the prompts to finish the retelling.
Additionally, E1 and E2 agree that generative images can promote users’ recall in the retelling, \zhenhui{while} sentence prompts are not necessary or even unhelpful.
% \textbf{DP2}) During each round of the repeated retelling stage, offer the generative images to help users recall the storyline. % by leveraging established connections. 
% 3. image-text alignment [highlight?]
% 3) Employ spatial proximity by placing highlighted text and images close together to facilitate users in making connections between images and text.
% 3) Employ spatial proximity by positioning highlighted text and images in close proximity, aiding users in effortlessly establishing connections between visual and textual content, and enhancing overall comprehension.

\textbf{DP3: To help users align the images and story content, \zhenhui{the system should} enable the users to select and enlarge an image while highlighting the related story sentence. }
As one of the twelve multimedia instructional principles \cite{mayer2002multimedia}, the spatial contiguity principle suggests that users could \zhenhui{be more focused on the learning tasks} when related text and image are \zhenhui{visually} close to each other. % learn more focused 
With the \name{} prototype, five learners in the \zhenhui{evaluation} study also suggest that the images should align with the story content in a more clear way. %within-subject
``I have to consciously remind myself to combine the images to understand the text. Showing all the images simultaneously and fixedly makes it difficult to focus on text and images at the same time'' (L1). 
% Thus, both 
Our English teachers help us identify the proper design to visually align the images and story content.
``Interaction design for displaying images should strike a balance between individual images and the overall narrative. \zhenhui{We could use an image slider that helps learners focus on one image at a time while having an overview of the image sequence}'' (E1). %, such as components that include both thumbnails and the current slide'' (E1).
``Highlighting the corresponding story sentence when the users enlarge one of the images \zhenhui{could be an} intuitive way'' (E2).
% \textbf{DP3}) To help users align the images and story content, enable the users to select and enlarge an image while highlighting the related story sentence. % When users comprehend and review the learning materials, show the coherent story and align the generative images with the corresponding sentence to help users establish the association between text and images. 
% build mental representations from text and visual elements.
% 4. transcription
% 4) Provide speech transcription capabilities to support students in reviewing, refining, and reflecting upon their retelling content. 
 % on the semantic correctness of the use of target vocabulary in the content of the retelling.

\textbf{DP4: In the story retelling stage, \zhenhui{the system should} provide a speech transcription function to record the users’ retelling content and help them keep track of their progress. }
As a similar feature with previous retelling-based English learning systems like \textit{CoSpeak} \cite{10.1145/3462204.3481750} and \textit{EnglishBot} \cite{ruan2021englishbot}, our participants in the formative study express their wish to check what they just spoke in the retelling exercise. 
``I want to see what I have said so far when retelling, which can help me organize what I will say next'' (S2).
In general, all ESL learners in the prototype evaluation and both teachers favor the component of speech transcription.
``With speech transcription, I could pay attention to the pronunciations when speaking'' (L14).
% \textbf{DP4}) In the story retelling stage, provide a speech transcription function to record their retelling content and allow them to edit the transcript to correct the wrong part. 
 % support students in refining and reflecting on the semantic correctness of the usage of target words in their retelling content. 
% 5. feedback [with images and coherent story]
% E1 suggests that to make users focus on the incorrect usage of vocabulary as well as prevent overloading, it is better to make users click on the incorrect vocabulary for more detailed feedback when they need it. 

\textbf{DP5: After each round of repeated retelling, to help learners review their performance, \zhenhui{the system should} offer feedback on the incorrect usage of target words, together with the story and generative images. }
% \textbf{DP5}) After each round of repeated retelling, to help learners review their performance, offer feedback on the incorrect usage of target words together with the story and generative images. %  and fluency of spoken language 
% When users finish retelling, offer adaptive feedback to assist users in assessing their retelling performance and focusing on the incorrectly used target words for timely correction and review. 
Providing feedback on users' task performance is a common and effective feature in learning support tools like \textit{ArgueTutor} \cite{wambsganss2021arguetutor} and \textit{EnglishBot} \cite{ruan2021englishbot}.
Five participants in the formative study suggest that they want to get feedback on their performance in practice, \eg about the correctness of words’ expressions. %of seven
``It \zhenhui{will be} better \zhenhui{if} the system could indicate whether I was using the target word correctly in my retelling practice, which can help me make progress in the next retelling'' (S1).
More importantly, eight participants in the evaluation study account the feedback from \name{} prototype for \zhenhui{their} perceived improvement \zhenhui{in the learning outcome}. % in the feedback from \name{} prototype. %within-subject
``Unlike the baseline system, \name{} tells me how well I did in the last exercise, which helps me recheck the target words’ meanings and make progress in the next \zhenhui{round of} retelling'' (L11).
\qiaoyi{
\chenqy{
E1 and E2 concur on the role of assessing the correctness of semantic usage through similarity measures and agree that it enables learners to verify the accuracy of their semantic expressions.
}
% E1 and E2 are in agreement about the function of judging the correctness of semantic use through similarity and believe that it can help learners to check whether semantically correct expressions are used.
}
However, \zhenhui{in} the \zhenhui{evaluation} study, seven learners suggest that \name{} would better present the feedback together with the story and images, so that they can better review their performance in the \zhenhui{current} round of repeated retelling before proceeding to the next round. %according to
``I hope to review the story and images again before starting the next round of retelling since it helps me fill in some of the details for the retelling'' (L13). 
E1 and E2 also agree that the review of the story and images between two rounds of repeated retelling is helpful. 
}

\peng{
\section{\name{} System Design and Implementation}
\label{sec:system_implementation}

Based on the identified design principles and proposed story text-to-image generation workflow in the last section, we develop \name{} to facilitate vocabulary learners in story retelling practices.
We develop \name{} as a web app that can be easily accessed by learners on their computers.
As shown in \autoref{fig:1}, in the structured procedure of a repeated retelling practice,  
\name{} provides users with generative images aligned to the story sentences (DP3) to assist story comprehension (DP1) and repeated retelling (DP2), speech transcription during repeated retelling (DP4), and adaptive feedback after each round of retelling practice (DP5). 
\zhenhui{We describe how vocabulary learners can use \name{} in a story retelling practice as follows.}

\zhenhui{
\textbf{Story comprehension.}
At the beginning of a story retelling practice, users need to first acquaint themselves with the target words' meanings and read the story that contains the target words (\autoref{fig:2}-A). 
At this stage, they can look up the bilingual definitions and pronunciation of each target word in the left part of the interface (A1). 
They can read the story with the targeted words marked in bold (A2). 
Meanwhile, users can click the ``Play'' button to listen to the audio of the story and click the ``Translation'' icon to check its Chinese meanings (A2).
Furthermore, users can click each image preview to switch the enlarged image (A3). 
Such a ``sliding'' interaction design with the images could help users focus on processing one image at a time and could be engaging \cite{sundar2014user}. 
% the generative images are shown with an interactive transition of ``slide'' (A3) to provide a dynamic and engaging user experience \cite{sundar2014user}. 
Users can also see a highlighted sentence in the story (A2) that corresponding to the enlarged image. 
Such a design follows the spatial contiguity principle, which states that users can learn better when related text and image are close to each other \cite{mayer2002multimedia}. 
}

\begin{figure}[htbp]
    \centering
    \includegraphics[width=14cm]{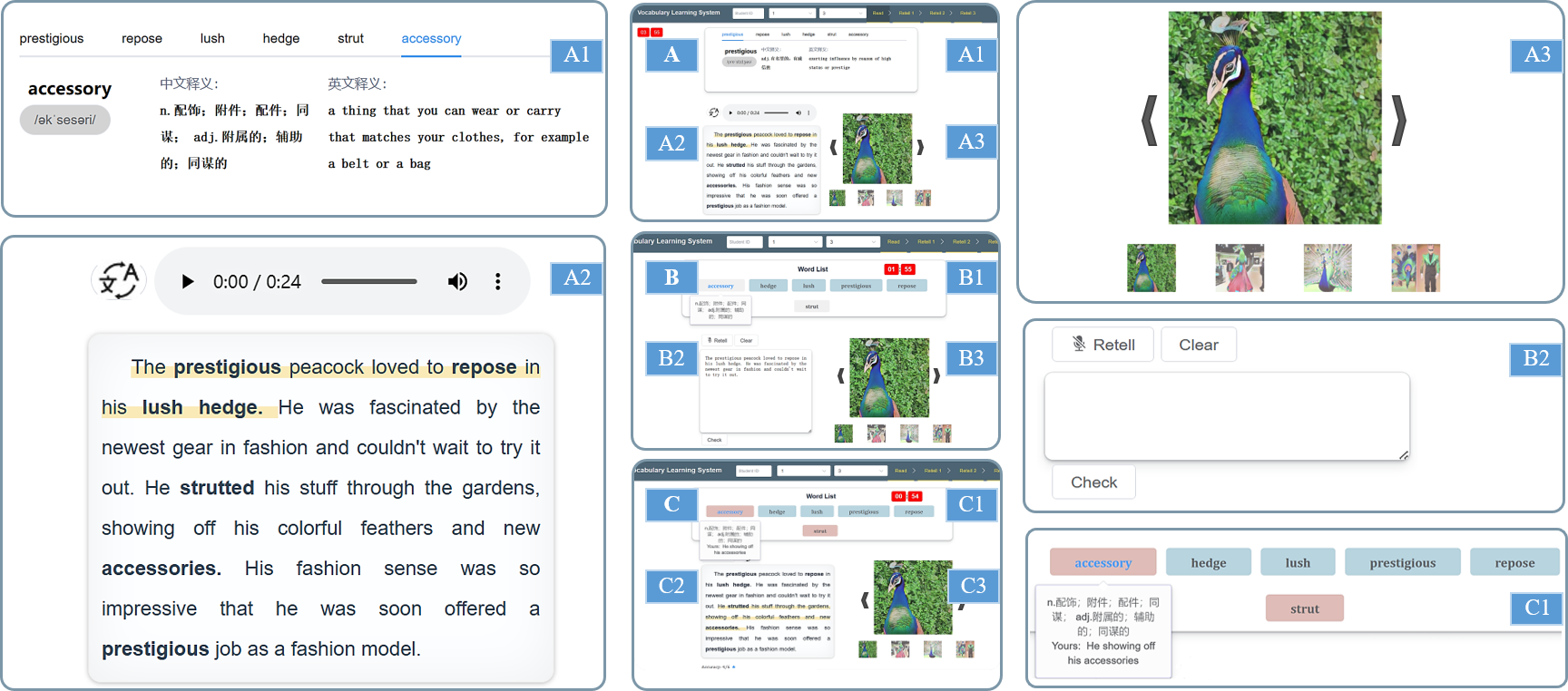}
    \Description[Screenshot]{Screenshot of the RetAssist user interface, showing different stages of interaction.}
    \caption{
    User interface design of \name{}. (A) In the story comprehension stage, users can 1) check the target words’ meanings, 2) read the story, and 3) see the relevant images for each story sentence. (B) In the repeated retelling stage, users can retell the story with 1) the target words, 2) the retelling transcription, and 3) the generative images. (C) After each round of retelling, users can check feedback on their performance and review 1) the target words with incorrect marks, 2) the story, and 3) the generative images. 
    % User interface design for story comprehfension. A: navigation window entering to Read and Retell. B: story comprehension module with dictionary module, story module and picture module. B1: dictionary module. B2: story module. B3: picture module corresponding to the story.
    }\label{fig:2}
\end{figure}

\textbf{Repeated retelling.}
% \subsubsection{Retell the story with target words. }
% The story retelling process is displayed in the interface (Figure \ref{fig:2}B). 
After comprehending the target words and associated story, users can click ``Retell'' in the upper menu bar to proceed to the repeated retelling stage (\autoref{fig:2}-B).
They need to complete three rounds of retelling practices within decreasing time limits, \eg 120, 90, and 60 seconds \zhenhui{based on our trials in the formative study and evaluation study of \name{} prototype}. %based on our formative study. 
The design of decreasing time limits \zhenhui{in the learning practices} could help users develop language fluency \cite{nation2007four}.  
Users can click ``Retell'' to start each round of retelling trials (B2). 
As they speak, \name{} will transcribe their speech in real-time using Chrome's speech recognition API \cite{ren2019fastspeech}. 
% During story retelling, adaptive prompts are provided in three levels---words, images. 
During each round of repeated retelling,
users can access the pronunciation and definition of target words in the word list any time they want (B1).
The background \zhenhui{color} of the word will turn ``blue'' when \name{} detects that the user speaks it. 
% The aid of visual elements (such as images) can help learners to understand the story correctly \cite{oktarina2020effectiveness, filippatou1996pictures}. 
% CTML considers that visual information may improve the recall of English target words on a later date as the learners are able to recall them based not only on the native verbal information but also on the shared visual information \cite{paivio1980dual}.
% CTML (Cognitive Theory of Multimedia Learning) is that supporting the dual-coding of visual information with L1-L2 verbal information during a L2 vocabulary learning phase may improve the recall of L2 words on a later date as the learners are able to recall the L2 words based not only on the L1 verbal information but also on the shared visual information \cite{paivio1980dual}. 
\zhenhui{Meanwhile, users can switch the image slider and click each generative image to enlarge it whenever they want (B3).
Users can stop the current round of story retelling by clicking ``Retelling'' again. 
Then, they can edit the transcribed sentences to correct speech recognition errors in the text box if they want (B2).
}
\phantomsection
\textbf{Review after each round of repeated retelling.}
\label{subsubsec:check} 
% After a round of story retelling, 
% Users can stop the current round of story retelling by clicking ``Retelling'' again. 
% Then, they can edit the transcribed sentences to correct speech recognition errors in the text box if they want (B2). 
% users can click ``Retelling'' again to stop this round of retelling practice and edit the transcribed sentences in the text box if they want (B4). 
% \cqyrevise{
% They can click ``check'' to view adaptive feedback regarding the accuracy of the target word usage in this round and review the story material with generative images (C). 
% % They can click ``check'' to view adaptive feedback regarding the accuracy of the target word usage in this round and review the story material (C). 
% }
\zhenhui{When users finish one round of repeated retelling, they can click the ``Check'' button (\autoref{fig:2}-B2) to view \name{}'s feedback on the their performance and review the story material with generative images (C).
% Following repeated retelling, users can make further modifications using the provided text box (Figure \ref{fig:2}B4) and click "feedback" to generate adaptive feedback on the retelling, including accuracy, fluency and review.
Users can check which target words have been correctly contextualized (marked in blue in C1) and which words are not used or incorrectly used in the repeated retelling (marked in red).  
% The accuracy measures the number of target words that have been correctly contextualized (marked in blue). 
% The non-used or incorrectly used words will be marked in red. 
% and the background of the words (C1) would turn red if the users incorrectly use it. 
They can click each red word to view its meanings and the associated sentence that the user spoke. %retells. %wrong % has retold.
Users can also read the story with the highlighted sentences that contain the target words they incorrectly use (C2). 
Meanwhile, they can check the associated generative images (C3). 
Users can click the ``Retell' button in the upper bar to start the next round of repeated retelling. 
}
% The review section also provides the story and highlights the sentences which the user incorrectly used the target word.
% definitions of incorrectly used target words and a comparison of the original and users' spoken sentences.
%and the user's incorrect usage. %(Figure \ref{fig:2}C). 
% After completing self-checking and review, users can click next ``Retell'' to proceed to the next round of repeated retelling or choose another story to start a new story retelling practice. 

\chen{
% We use semantic similarity to judge whether the user correctly uses the target words. % do not remember the word meaning
% [We calculate the semantic similarity between two sentences by ...]
% We assess the accuracy by semantic similarity. 
\chenqy{
% \zhenhui{We use semantic similarity to judge whether the user correctly uses the target words.}
We use semantic similarity to judge whether the user correctly uses the target words, inspired by the study of Cao \etal \cite{cao2022semmt}, which verifies the correctness of machine translation by checking semantic similarity between the original and the translated sentences.
}
Specifically, we consider a target word is not correctly used if the spoken sentence that should contain this word is semantically different from the original story sentence that contains this word \cite{rubenstein1965contextual}. 
\cqyrevise{
We calculate the semantic similarity (ranging from 0 to 1) between the expression of each target word in the story and that in the users’ retelling. 
First, we identify the sentence containing each target word in the user’s retelling and calculate their sentence embeddings by Sentence-BERT \cite{reimers2019sentence}. 
Then, we compute the cosine similarity (ranging from 0 to 1) between this identified sentence and the corresponding sentence from the original story. 
If the user mentions the target word in multiple sentences, the similarity is recorded as the maximum of similarity between the multiple sentences and the corresponding sentence from the original story. 
If the user does not mention the target word, the similarity is recorded as 0. 
% we calculate the semantic similarity (ranging from 0 to 1) between the expression of each target word in the story and that in the users' retelling. 
% First, we identify the sentence containing each target word in the user's retelling and calculate their sentence embeddings by Sentence-BERT \cite{reimers2019sentence}. 
% We then compute the cosine similarity (ranging from 0 to 1) between this identified sentence and the corresponding sentence from the original story.
% If the user does not mention the target word, the similarity is recorded as 0. 
}
To decide the thresholds of similarity scores that differentiate the correct and incorrect use of target words, three authors mark the correctness (\ie correct or incorrect) of the word usage in each sentence of the recorded retelled content of the participants in our formative study. 
\cqyrevise{After marking, we obtain two sets of similarities separately representing the correct use of word meanings and the incorrect use of word meanings by calculating the semantic similarities between story sentences and spoken sentences. }  
Finally, the threshold is determined to be 0.7 based on the ROC curve for different similarity scores \cite{fawcett2006introduction}. 
}

}
\peng{
\section{User Study}

% We conducted a user study to explore the experience and effectiveness of our interactive vocabulary learning system, \name{}, in assisting with learners' vocabulary acquisition through story retelling, as well as to investigate how the assistance provided by \name{} affects learning outcomes. 
To evaluate how the generative images in \name{} impact users' vocabulary learning outcome and experience in the story retelling practices, we conduct a within-subjects (\name{} vs. baseline) study with 24 ESL (English-as-the-Second-Language) university students in China. 
% The study involves 18 university students and used a within-subjects design to evaluate two conditions for vocabulary-focused story retelling practice with and without intelligent assistance.
% In the case with intelligent assistance, users read and comprehend the story containing the target vocabulary with visual aids provided by \name{} (Figure \ref{fig:1}), and are required to complete repeated retelling of the story within decreasing time limits with the assistance of visual aids, in-situ prompts and adaptive feedback(Figure \ref{fig:2}). 
% In the case without intelligent assistance, \ie in the baseline system condition, users learn English vocabulary by observing stories containing the target vocabulary (Figure \ref{fig:3}), and repeatedly retell the story in on their own within reducing time limits but without any intelligent assistance and only to proofread their content of retelling after each practice (Figure \ref{fig:4}).
Our research questions are: 
% \textbf{RQ1.} How do the intelligent assistance provided by the interactive vocabulary learning system affect users’ a) engagement and enjoyment and b) retelling performance in the vocabulary-focused story retelling practice?
% \begin{itemize}
% \item[$\bullet$] 

\textbf{RQ1.} How would \name{}\zhenhui{'s generative images} affect users’ learning outcomes regarding the retention and verbal expression of target words in their story retelling practices? % vocabulary-focused 

\textbf{RQ2.} How would \name{}\zhenhui{'s generative images} affect users’ a) learning experience and b) behaviors in their story retelling practices?  %learning process such as retelling vocabulary-focused 
% \cqy{Workload}
% \item[$\bullet$] 
% \textbf{RQ2.} How do the intelligent assistance provided by the interactive vocabulary learning system affect users’ learning outcome in the vocabulary-focused story retelling practice?
% \item[$\bullet$] 

% \textbf{RQ3.} How do users perceive such interactive vocabulary learning system providing assistance in the vocabulary-focused story retelling practice?
\textbf{RQ3.} How would users perceive the usefulness of \name{}'s generative images in their story retelling practices? %vocabulary-focused
% \end{itemize} that provides generative images

\begin{figure}[htbp]
    \centering
    \includegraphics[width=14cm]{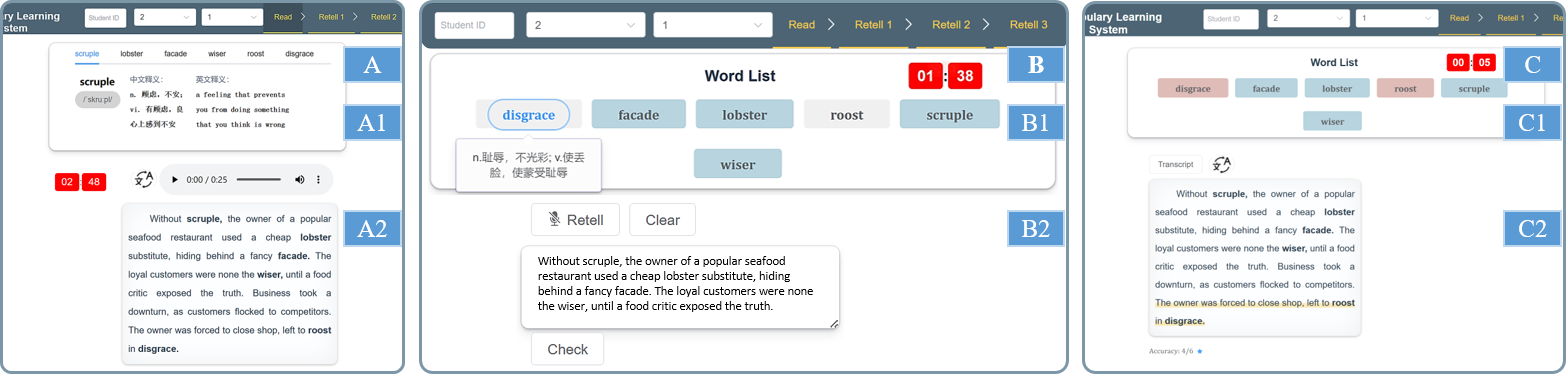}
    \Description[Screenshot]{Screenshot of the baseline system user interface, showing different stages of interaction.}
    \caption{
User interface design of the baseline system. (A) In the story comprehension stage, users can 1) check the target words’ meanings, and 2) read the story. (B) In the repeated retelling stage, users can retell the story with 1) the target words, and 2) the retelling transcription. (C) After each round of retelling, users can check feedback on their performance and review 1) the target words with incorrect marks, and 2) the story. \zhenhui{The baseline system differs from \name{} in that it does not provide generative images.}  
    }\label{fig:3}
\end{figure}
\subsection{The Baseline System}
\cqyrevise{
The baseline system (\autoref{fig:3}) supports the same user workflow (\autoref{fig:1}) in story retelling practices as \name{}. However, it does not offer generative images during both the story comprehension stage and the repeated retelling stage. The baseline system simulates the scenario in which the user is required to learn target words via story retelling practices without generative images. Specifically, the baseline system offers the word list (\autoref{fig:3}-A1) and story (\autoref{fig:3}-A2) in the story comprehension stage, and it provides decreasing time limits as well as word list (\autoref{fig:3}-B1) and speech transcript (\autoref{fig:3}-B2) in users’ three rounds of retelling practices. Also, users can check adaptive feedback regarding the accuracy of the target word usage (\autoref{fig:3}-C1) in such rounds and review the story text (\autoref{fig:3}-C2). 
}
\qiaoyi{
Such a baseline system satisfies all design principles without the involvement of generative images, specifically referring to DP4 and DP5. In summary, the only difference between \name{} and the baseline system lies in the incorporation or exclusion of generative images, while all other functionalities are present in both conditions to meet users' demands.
}

\subsection{Participants}
We recruit 24 undergraduate students (P1-24, 15 females, 9 males, mean age: 20 (SD = 1.67)) from a university in mainland China via \zhenhui{a post in the social media.}%word-of-mouth. %, and they are all typical ESL learners.  
% Over 11 distinct college majors were represented, including computer science, historiography, philosophy, physics, finance, literature, and international relations.
They major in various domains such as Computer Science, Historiography, Philosophy, Physics, Finance, Literature, and International Relations. 
Twenty-three participants have passed the national English exam CET-4 in China, with an average score of 575 (SD = 48.04) \footnote{710 is the full mark of both CET-4 and CET-6, and 425 is the minimum score to pass the exams.}. 
Seventeen participants additionally have passed a higher-level national exam CET-6 in China (Mean score: 523 (SD = 48.47)).
None of our participants have taken the IELTS exam. % yet. %, and three are preparing for it.
% \zh{None of them have taken the international English examine IELTS?}. 
% while others have not participated in it. 
However, they exhibit a strong interest in learning their unknown IELTS vocabulary via the story telling practices (M = 5.75, SD = 1.05; 1 - not interested at all, 7 - very interested). %, with an average rating of 
% Meanwhile, participants self-assess their proficiency in IELTS vocabulary with a score of 3 (SD = 1.26; 1 - not proficient at all, 7 - very proficient), 
% In general, they report moderate amount of confidence in spoken English. 
% and they assess their confidence in English speaking at 4.5 (SD = 2.26; 1 - not confident at all, 7 - very confident).

\subsection{Procedure and Tasks}
\label{subsec:procedure}
\begin{figure*}[htbp]
    \centering
    \includegraphics[width=12cm]{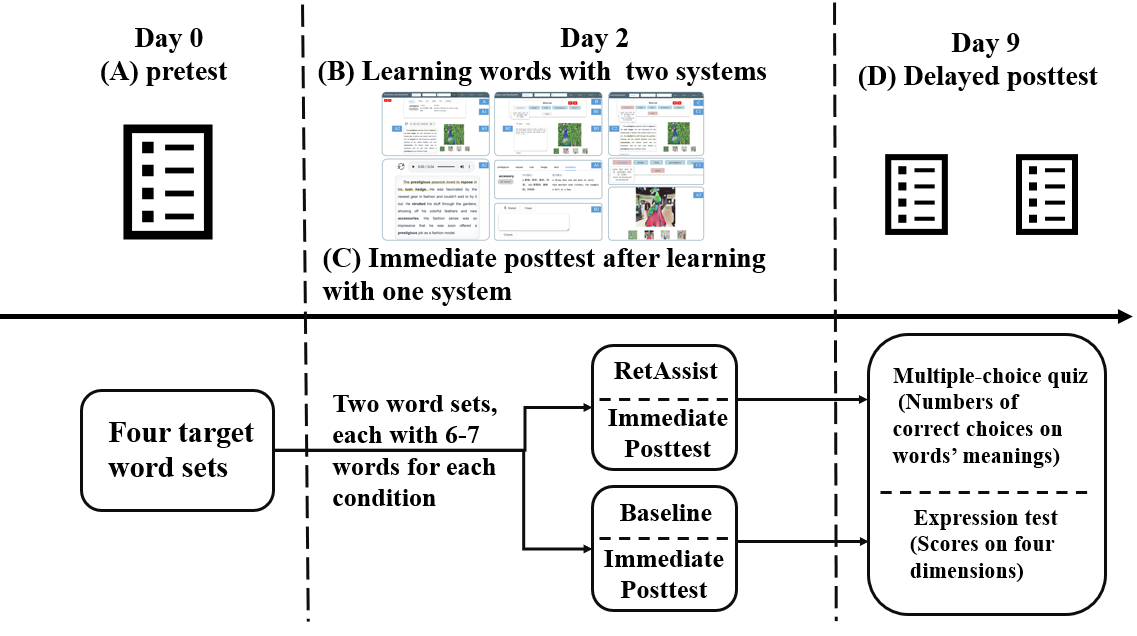}
    \Description[Diagram]{Diagram outlining the steps include pretest (Day 0), learning words with two systems and immediate posttest (Day 2), and delayed posttest (Day 9). }
    \caption{
Procedure of the within-subjects (\name{} vs. the baseline system) user study. In each task, participants learn two sets of target words with either the \name{} or the baseline system. 
}\label{fig:procedure} % based on their assigned order
\end{figure*}
% All participants were invited to our lab for on-site learning sessions to ensure stable Internet connections and minimize external disturbances. 
% We conduct the experiment with each participant offline. 
% We conduct an online experiment with each participant. 
\autoref{fig:procedure} shows the procedure and task of our user study conducted remotely. % We conduct user study remotely. 
% We set up our \name{} and baseline system in three laptop computers. 
% We assign participants to each computer with an experimenter, which enables us to run the experiment in parallel. 
% We assign 5-6 participants to run the experiment simultaneously.
Following \cite{arakawa2022vocabencounter,peng2023storyfier}, on Day 0, participants fill in a consent form and a background survey and take a vocabulary pretest. 
% The pretest consists of 26 (?) IELTS words that the first author selects and group into four sets, each with six or seven target words. 
\cqyrevise{
We randomly select 4 stories from the 20 prepared stories mentioned in \autoref{subsubsec:stories} as the learning materials for all participants, each containing six or seven target words.
% We randomly select 4 stories from the 20 prepared stories mentioned in \autoref{subsubsec:stories} as learning materials, each containing six or seven target words.  
}
\zhenhui{In total,} the pretest consists of the 26 target words that participants will learn in our learning sessions. 
% The first author selects and groups these words into four target word sets, each with six or seven target words. 
% Participants have two vocabulary learning sessions. 
% Referring to the procedure in \cite{arakawa2022vocabencounter}, the procedure of our user study consists of three stages (Figure \ref{fig:procedure}). 
% First, with consent of participants, we gather their background information and ask them to take a pretest so as to identify the target words they have known. 
For each target word in the pretest, participants are required to select one option from five choices, including one that gives the correct meanings of the word in Chinese, three distractors, and an ``I don't know'' option. 
% In the pretest, participants 
% % are presented with both word lists and 
% are required to select one option for each target word from five choices, including one correct Chinese meaning, three distractors, and the ``I don't know'' option. 
% We confirm that all participants select the incorrect option at least 13 times, indicating that they do not know the correct meanings of more than half of our prepared target words. 
\chen{
According to the results of the pretest, the average number of \zhenhui{correctly chosen} options among 24 participants is 8.62. \zhenhui{In other words, on average, participants do not know the meanings of 17.38 words prior to the learning sessions.} %(26 words in total). 
}
% \chen{We recruit participants with less than 50\% accuracy, indicating that they do not know the correct meanings of half of our prepared target words.}
% had at least half of the target words they have not known. 
% Besides, 
We inform them not to learn the words that appear in the pretest before the learning sessions. 

On Day 2, participants first watch our pre-recorded video that describes the learning task and introduces the interfaces of \name{} and baseline systems with blind names. 
They then use their laptops to log in to our systems. 
Each participant has two learning sessions. 
In each session, participants have two story retelling practices with either \name{} or baseline system to learn two target word sets that they encountered in the pretest. 
\cqyrevise{
Based on the pilot study with two participants, we allocate 30 minutes for each learning session. 
% Based on the pilot study with two participants, we recommend them spend around 30 minutes in each learning session and inform them that they can finish early or spend more time if desired. 
}
% For each condition, participants were allotted 30 minutes to learn two sets of vocabulary and were informed that they could finish early or spend more time if desired. 
% After each learning task, participants are require to conduct an immediate posttest that examines their retention of target words via multiple-choice questions and assesses their expression of target words via retelling the story without the help of generative images. 
After each learning session, participants rate their engagement, enjoyment, task workload, and perceptions of the system in a questionnaire. 
\chen{In the questionnaire, we also ask them to write down responses to some short questions so as to make sense of the ratings. }
% Upon the conclusion of each condition, the participants provided ratings pertaining to their level of engagement, enjoyment, perceived learning performance, and acceptance of the system used within the respective condition. 
Additionally, they need to conduct an immediate posttest that examines their learning outcome on remembering the target words' meanings and being able to verbally use them to retell a story. %retention of target words via multiple-choice questions and assesses their expression of target words via retelling the story without seeing the textual story and/or generative images. %examines their retention and expression of target words. 
% Upon completion of four practices with both systems, we ask participants' opinions and suggestions on our vocabulary learning support systems. 
% Subsequently, participants completed an immediate vocabulary test, including multiple-choice questions for vocabulary comprehension and a vocabulary expression test based on a Chinese script. 
% After completing both conditions, we solicited the participants' opinions on our interactive vocabulary learning system and any suggestions for improvement. 
\chen{Upon completion of two learning sessions, participants fill in a questionnaire that asks them to write down their preferences on the interfaces, comments on the generative images, and suggestions for improving \name{}. }
% Upon completion of two learning sessions, we further ask for their preferences on the interfaces, comments on the generative images, and suggestions for improving \name{}. 
We counterbalance the order of the used systems and word sets using Latin Square, \ie six participants experience ``set 1 and 2 with \name{} --> set 3 and 4 with Baseline'', six ``set 1 and 2 with Baseline --> set 3 and 4 with \name{}'', six ``set 3 and 4 with \name{} --> set 1 and 2 with Baseline'', and the rest six ``set 3 and 4 with Baseline --> set 1 and 2 with \name{}''. % that the two systems interfaces are used and the word sets to learn with a particular system.

On Day 9, they conduct a delayed posttest that has the same format as the immediate posttest to examine their retention and verbal expression of target words learned on Day 2. 
The procedure on Day 2 and Day 9 is video- and audio-recorded for further data analyses.
% Forty-eight hours after completing the experiments in both conditions, participants were required to take the vocabulary tests again. 
Overall, each participant spends approximately one hour and a half in our study and receives 80 RMB as compensation. 
}

\peng{
\subsection{Measurements}
\subsubsection{RQ1. Learning outcomes}
We measure participants’ vocabulary learning outcomes through performance on an immediate posttest right after each learning session and a delayed posttest one week later.
 % vocabulary tests administered after learning tasks. 
Specifically, both posttests include a multiple-choice quiz and an expressive test. % same as the pretest to measure participants' retention of target words' meanings and an expressive test to evaluate how well the participants master the expressions of target words. 
% Specifically, we similarly conducted immediate and delayed posttests as \cite{10.1145/3025453.3025779}, each including vocabulary multiple-choice questions and expressive tests. 
The multiple-choice quiz is the same as the pretest that requires users to select one of five options that is the correct Chinese meaning of the target word. 
% In the multiple-choice quiz, participants are required to select one option for each target word from five choices, including one correct Chinese meaning, three distractors, and the ``I don't know'' option. 
We calculate the number of correct answers to the multiple-choice questions to capture the learning outcome on the meanings of target words.
% whose meanings are correctly memorized after learning with \name{} and the baseline system. 
% In the vocabulary multiple-choice quiz, learners were required to complete vocabulary tests of the learned word groups. 
% Each vocabulary multiple-choice question included one correct meaning, three distractors, and an "I don't know this word" option. 
% Choosing the correct meaning earned one point, while choosing incorrectly resulted in no points.

In the expression test, participants need to verbally retell each story in their learning sessions based on the story synopsis in Chinese and the target word set. 
As suggested by our two English teachers in the design process, we choose to present the synopsis instead of presenting nothing or providing the full Chinese translation of the original story to balance the difficulty of the expression test. 
% learners are asked to retell a story based on a Chinese script. 
% The scoring criteria for this test have a focus on the target words and are adapted from the marking scheme of the IELTS speaking test \cite{cullen2012cambridge}.
We adapt the marking scheme of the IELTS speaking test \cite{cullen2012cambridge} but have a focus on the verbal expressions of target words. 
% include counting around target words and fluency. 
With confirmation from our two English teachers in the design process, for each expression test of two stories within a learning session, we capture: 

\begin{itemize}[leftmargin=*]
    \item \textbf{Number of target words used} (range 0 - 13 \footnote{In each learning session, participants learn vocabulary based on two stories. One contains 6 target words, and the other contains 7 target words. The maximum score for one learning session is therefore 6 + 7 = 13.} ). 
    \item \textbf{Number of target words pronounced correctly}, \ie the number of target words correctly pronounced. 
    \item \textbf{Number of target words used correctly}, \ie the number of target words that have been used semantically correctly. 
    \item \textbf{Fluency}, which is determined by the expression of individual clauses and the lag between sentences, ranging from 0 to 9 on a scale referenced to the IELTS marking scheme. 
\end{itemize}

Three authors of our research team first independently score six randomly selected audio samples, each consisting of two retelling stories in a learning session. 
They then meet and discuss together with one of our two English teachers (male, age: 29) to refine their rating scheme.  %in a Chinese University 
For example, to focus on the usage and expression of target words, the rating scheme excludes factors like the participants' volume of voice, intonation, or accent. 
The three authors then apply the rating scheme to all 192 (24 $\times$ 2 systems $\times$ 2 stories per system $\times$ 2 posttests) audio samples in a shuffled order. 
% They do not have access to the participant information and the used vocabulary learning system of the audio sample 
% For the three authors' scoring results, we re-rate the scores that are inconsistent in the first three dimensions to reach agreement, and 
For each dimension of the measured performance on the verbal expressions of target words, we average the three authors' scores (ICC = $0.939$) as the final score in each retelling story. %(intraclass correlation coefficient)
% We average the scores of fluency of to evaluate participants' learning performance on both systems based on the four stories learned by each participant (two stories for each system). 
For each of the first three dimensions, we add the scores of two stories within one learning session as user performance in verbally expressing target words learned in that session, %, we separately sum the counting of the first three dimensions 
while for the last dimension of fluency, we average the scores of the two stories as the final score. 
%selected at random, scoring them individually before engaging in a thorough discussion to establish a consensus on the rating criteria. The rating scheme excluded factors such as the user's voice, intonation, or accent. Subsequently, each annotator scored all 144 audio samples in the posttest independently; $ICC(intraclass correlation coefficient) = 0.838, p < 0.001$.
% By comparing the correctness rate of multiple choice questions and expressive scores on four scoring dimensions between the experimental conditions, we evaluated how \name{} contributed to outcomes of learners' comprehension and expression skills related to English vocabulary.

\subsubsection{RQ2. Learning process}
In each learning session with either \name{} or baseline system, we measure participants' engagement and enjoyment in the learning process using items adapted from \cite{wambsganss2021arguetutor,wu2020predicting}: ``I was absorbed in using this interface to learn vocabulary''  and ``It is enjoyable to learn vocabulary with this interface''. 
Besides, we measure the perceived task workload of learning sessions using items adapted from NASA Task Load Index \cite{hart2006nasa} (\eg ``I require much mental and perceptual activity such as thinking and remembering in the process of the story retelling practice''). 
In addition to the questionnaire data, we also measure how learners perform in each of the three rounds of repeated retelling in each practice. 
For each round of repeated retelling, we measure: 1) spent time, \ie the time period between clicking the ``Check'' and the ``Retell'' button in this round of repeated retelling; 2) performance in practice, \ie how well users can retell the story content, reflected on the semantic similarity between learners' retold content and original story (ranging from 0 to 1, detailed in \textbf{Review after each round of repeated retelling} in \autoref{subsubsec:check}).
% \begin{itemize}[leftmargin=*]
%     \item Spent time: we log the time period between clicking the ``Check'' and the ``Retell'' button in this round of repeated retelling.
%     \item Performance in practice: 
%     % we assess how well users can retell the story content by calculating the semantic similarity (ranging from 0 to 1) between the retelled content and the story. 
%     % Specifically, we convert ... [please describe how you measure it]
%     \chen{we assess how well users can retell the story content by calculating the semantic similarity (ranging from 0 to 1), detailed in \autoref{subsubsec:check}. }
%     % Specifically, we first compute sentence embeddings for all story sentences in the story. 
%     % Then, for each target word in the users' retelling transcription, we embed the sentence where it lies. 
%     % Calculating the cosine similarities between the sentence embedding in the retelling transcription and the sentence embeddings in the original story, we take the maximum cosine similarity as the semantic similarity of the target word. 
%     % Specifically, we first identify the sentence containing each target word in the user's retelling.
%     % Then we compute the semantic similarity between this identified sentence and the corresponding sentence from the original story.
%     % If the user does not mention the target word, the similarity is recorded as 0. 
% \end{itemize} 
For each learning session with two story retelling practices, we average the spent time in two practices as the mean time spent in one round of repeated retelling in that session. 
Similarly, we average the semantic similarity scores of two practices to reveal user performance in one round of repeated retelling in each learning session. 
% og the spent time per round of repeated retelling (three rounds in one practice) and ii) compute the semantic similarity (range: 0 - 1 \footnote{We use xxx to ... [please describe how we calculate it and the meaning of range]}) between participants' sentences in their speech transcript and the original story sentences that contain the same target words.  
%``I was absorbed in using this version of the system to learn vocabulary in the last session'' and "It is enjoyable to learn vocabulary with this version of the system in the last session”.  
% Moreover, we measure users' retelling performance (RQ1b) in the three-time retelling practices with \name{} 's log of the number of adaptive prompts provided and the number of correctly used target words in each retelling practice.
% The frequency of prompts triggered by users and the correct rate of target vocabulary usage 
% Also, we record the length of time for each of the three retelling practices performed by users for both experimental conditions.

\subsubsection{RQ3. Perceptions towards \name{}}
% 0.921, 0.830, 0.901
For each system interface, we adapt the technology acceptance model \cite{venkatesh2008technology, wambsganss2020adaptive} to measure the perceived \textit{usefulness} (four items, e.g., ``I find the vocabulary learning support system useful in my vocabulary learning process by story retelling''; Cronbach’s $\alpha$ = 0.921); \textit{easiness to use} (four items, e.g., ``My interaction with the vocabulary learning support system is clear and understandable''; $\alpha$ = 0.830); and \textit{intention to use} (two items, e.g., ``I intend to be a heavy user of the vocabulary learning support system when I want to learn vocabulary''; $\alpha$ = 0.901). 
We average the ratings of multiple questions as the final score for each aspect. All statements in the questionnaires are rated on a standard 7-point Likert Scale, with 1 - strongly disagree and 7 - strongly agree. 
}

\begin{figure*}[htbp]
    \centering
    \begin{minipage}[t]{0.32\textwidth}
        \centering
        \includegraphics[width=4.9cm]{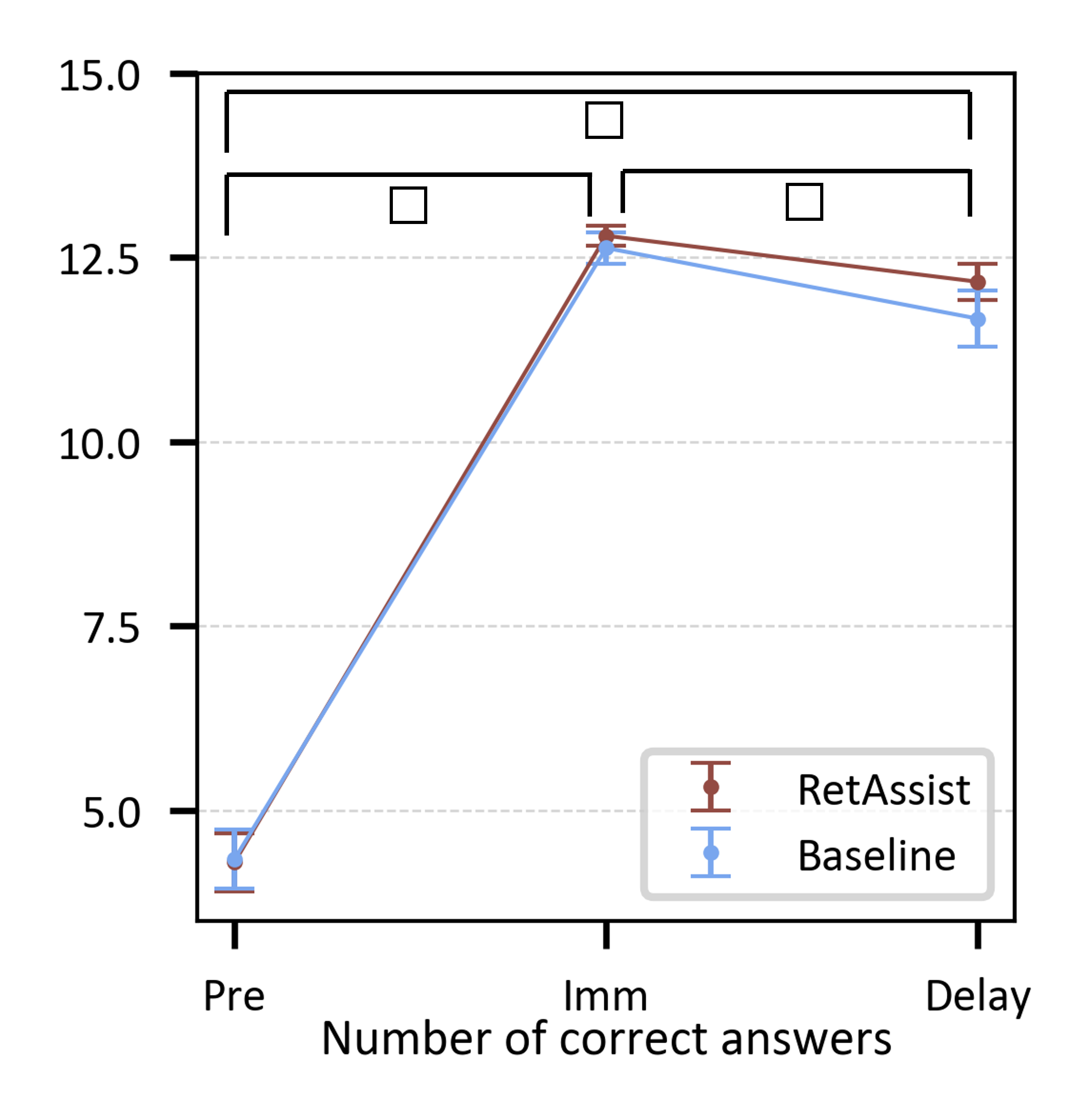}
        \Description[Line graph]{Line graph indicating significant differences between 1) pretest and immediate posttest, 2) pretest and delayed posttest, 3) immediate posttest and delayed posttest. }
        \vspace{-1cm}
        \caption{
     RQ1 results regarding the number of correct choices on target words’ meanings. \qiaoyi{$\square: p < .05$ for time factor (pretest vs. immediate posttest vs. delayed posttest) using repeated measures ANOVA with Bonferroni post-hoc test. }
    }\label{fig:learing_outcoming1}
    \end{minipage}%
    \hspace{0.2cm}
    \begin{minipage}[t]{0.32\textwidth}
        \centering
        \includegraphics[width=4.75cm]{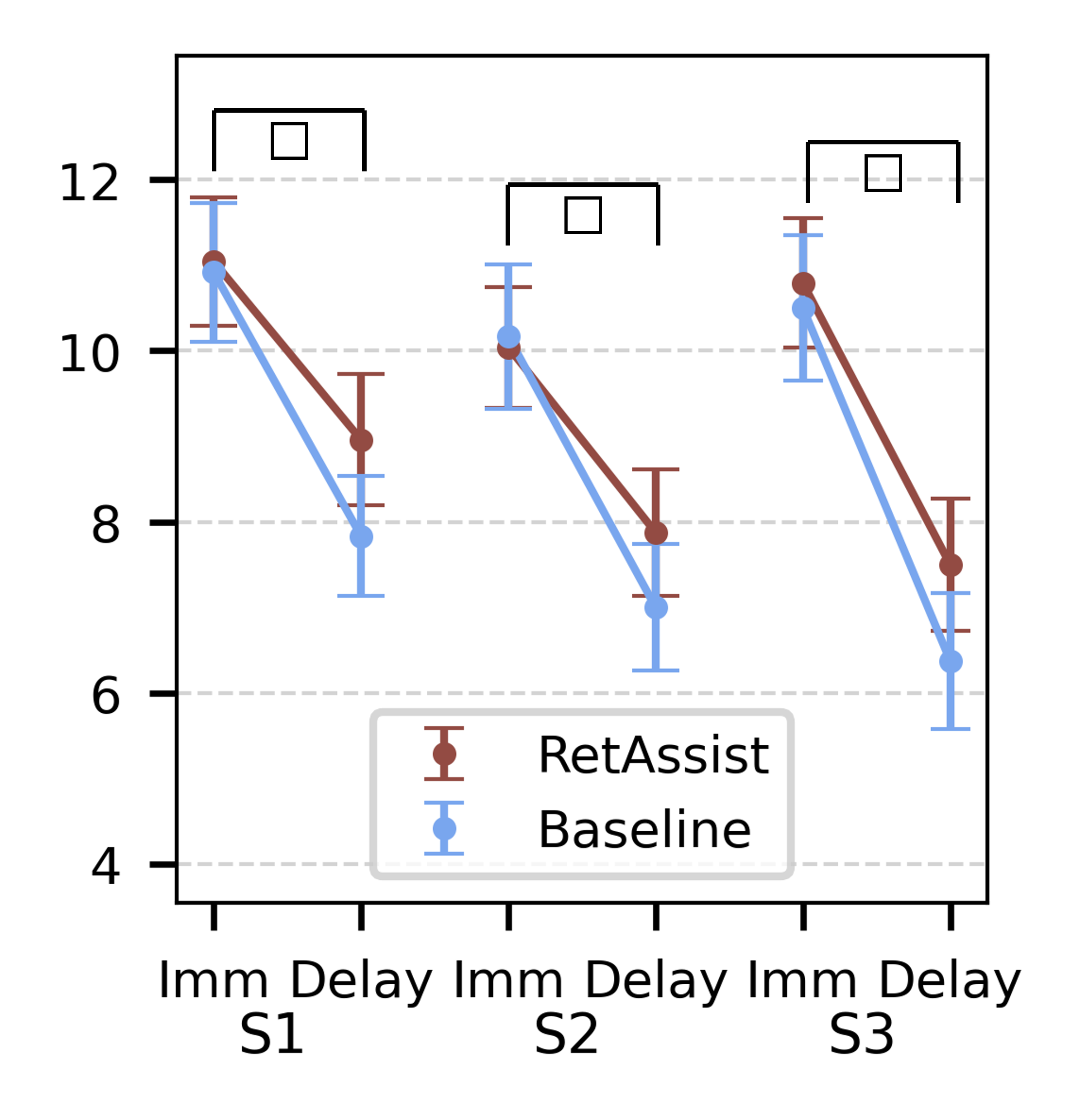}
        \Description[Line graph]{Line graph indicating significant differences between immediate posttest and delayed posttest in S1, S2 and S3. }
        \vspace{-1cm}
        \caption{
    RQ1 results regarding the number of target words used in expression (S1), the number of target words pronounced correctly in expression (S2), and the number of target words used correctly in expression (S3). \qiaoyi{$\square: p < .05$ for time factor (immediate posttest vs. delayed posttest) using repeated measures ANOVA. }
    }\label{fig:learing_outcoming2}
    \end{minipage}%
    \hspace{0.1cm}
    \begin{minipage}[t]{0.32\textwidth}
        \centering
        \includegraphics[width=4.75cm]{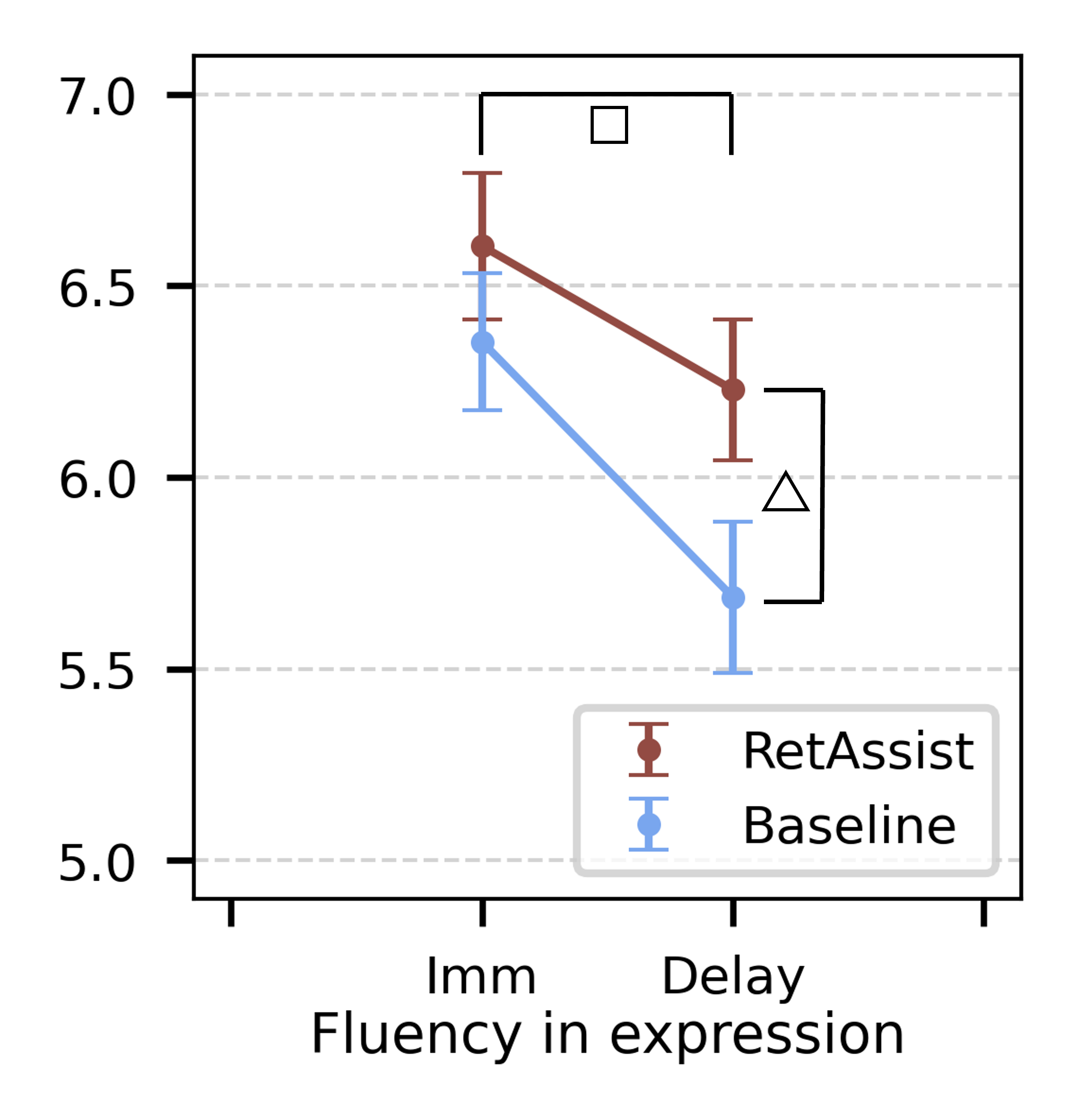}
        \Description[Line graph]{Line graph indicating a significant difference between RetAssist and Baseline as well as a significant difference between immediate posttest and delayed posttest. }
        \vspace{-1cm}
        \caption{
    RQ1 results regarding the fluency in expression. \qiaoyi{$\square: p < .05$ for time factor (immediate posttest vs. delayed posttest), $\triangle: p < .05$ for system factor (\name{} vs. Baseline) using repeated measures ANOVA. }
    % \qiaoyi{$*: p<0.05, ***: p<0.001$ using paired-sample t-tests. }
    }\label{fig:learing_outcoming3}
    \end{minipage}%
\end{figure*}

\peng{
\section{Analyses and Results}
\qiaoyi{
\chenqy{
For the rated items, we first conduct a set of mixed ANOVA tests to check whether the order of system usage or the learned word sets associated with the systems affected our results (order and word sets as between-subjects, systems as within-subjects). The results indicate that neither the main effects of the order and word sets nor their interaction with the systems are significant.
}
For the measurements for RQ1, we perform two-way (time and system) repeated measures ANOVA to account for the dependencies in time. 
% Specifically, for the number of correct choices on target words' meanings, we run a two-way mixed ANOVA (time as within-subject and gender as between-subject). 
As for the measurements for RQ2 and RQ3, we perform Shapiro-Wilk normality tests before running all the paired samples t-tests. 
If the hypothesis that the data satisfies a normal distribution is rejected, we use paired samples Wilcoxon signed rank tests instead. 
% and also performed tests for homogeneity of variance for all the ANOVA analyses.
% The assumptions on each rated item's normal distribution are confirmed based on the Shapiro-Wilk tests. 
% We also confirm that the orders of using systems and assigned learning materials do not significantly impact each item based on a set of mixed-design repeated measures ANOVA (system as within-subjects, order and learning material as between-subjects). 
% Therefore, the following results of statistic tests are based on the paired-sample t-tests. 
As a result, for the spent time and performance in each round of repeated retelling, we perform paired-sample t-tests to compare the \name{} and the baseline system. 
% For the ratings on the participants' scored learning outcome in meaning retention, self-reported learning experience, and perceptions toward systems, 
For the rest measures, we perform paired samples Wilcoxon signed rank tests. % to compare the \name{} and the baseline system. 
Additionally, two authors conduct open coding on participants' comments and suggestions on both vocabulary learning systems. 
They have multiple rounds of discussions and finally reach an agreement on the codes, which are incorporated into the following result presentation. 
% Since we have evenly distributed the order of using the two systems and the stories to be learned with each system, in the following repeated measured t-tests, we treat the system condition (\name{} vs. Traditional Read-and-Retell) as the only within-subjects variable. 
}

\subsection{Learning Outcomes (RQ1)}
\subsubsection{Multiple-choice Quiz}
As shown in \autoref{fig:learing_outcoming1}, participants demonstrate comparable performance with \name{} $(M = 12.792, SD = 0.644)$ and baseline system $(M = 12.625, SD = 1.033)$ regarding the number of correct answers to the multiple-choice questions in the immediate posttest.  % with \name{} $(M = 12.792, SD = 0.644)$ exhibiting a slight advantage in the number of correct answers compared to the baseline system $(M = 12.625, SD = 1.033)$ but no significance. 
% \qiaoyi{
% However, in the delayed posttest, there is a significant difference in the number of correct answers between the \name{} $(M = 12.167, SD = 1.179)$ and baseline system $(M = 11.667, SD = 1.863)$; $p < 0.05 (z = 2.23, \text{Cohen's d} = 0.314)$. 
% }
\qiaoyi{
In the delayed posttest, participants have better performance on average with \name{} $(M = 12.167, SD = 1.179)$ than baseline system $(M = 11.667, SD = 1.863)$ regarding the number of correct answers. 
The results of repeated measures ANOVA indicate that neither the system factor (\name{} and Baseline) nor its interaction with the time factor (pretest vs. immediate posttest vs. delayed posttest) significantly affects participants' performance in the multiple-choice quiz ($p > 0.05$). 
However, the time factor has significant effects on participants' performance in the multiple-choice quiz ($p < 0.001, F = 596.792, \eta^2=0.912$), and the results of the Bonferroni post-hoc test ensure the significant difference among the three quizzes (pretest vs. immediate posttest: $p<0.001$; pretest vs. delayed posttest: $p<0.001$; immediate posttest vs. delayed posttest: $p<0.05$). 
% Participants who utilize the \name{} exhibit significantly superior retention of words' meaning. 
% This result indicates that \textbf{\name{}\zhenhui{'s generative images} can improves the long-term retention rate of target words' meanings compared to the baseline system. }
}
% The results indicate that the image feature exerts a notable influence on the long-term retention of vocabulary.

\subsubsection{Expression Test}
As shown in \autoref{fig:learing_outcoming2}, participants generally perform well in using the target words, pronouncing them correctly, and using them correctly in the immediate expression posttest after the learning session with either \name{} or baseline system; $M > 10$ and $ p > 0.05$ in all the three dimensions. 
This finding suggests that the story retelling practice, either with or without the involvement of generative images, is an effective approach to learning the verbal expression of target words in the short term.
% and \autoref{fig:learing_outcoming3}, participants generally perform better in expressing target words learned with \name{} than those learned with the baseline system. 
% In the immediate posttest, neither the \name{} nor the baseline system affect the first three scoring dimensions.
In the delayed posttest after one week of the learning sessions, the user performance with both systems naturally decreases compared to that in the immediate posttest.
\qiaoyi{
% However, 
We find that participants are able to use more target words learned with \name{} $(M = 8.96, SD = 3.77$ and use them correctly $(M = 7.5, SD =3.77)$ 
compared to the baseline system (use target words: $M = 7.83, SD = 3.45$, use them correctly: $M = 6.375, SD = 3.89$) in average. 
The average number of correctly pronounced target words is also higher in the learning session with \name{} ($M = 7.875, SD = 3.61$) than that in the session with baseline system ($M = 6.375, SD = 3.89$). 
With the repeated measures ANOVA, we find that neither the system factor (\name{} and Baseline) nor its interaction with the time factor (immediate posttest vs. delayed posttest) significantly affects the number of target words used (S1), the number of target words pronounced correctly (s2), and the number of target words used correctly (S3) in expression ($p > 0.05$). 
Among the expression measurements of S1 - S3, the time factor has significant effects (S1: $p < 0.001, F = 49.127, \eta^2=0.699$; S2: $p < 0.001, F = 43.381, \eta^2=0.712$; S3: $p < 0.001, F = 76.741, \eta^2=0.827$).
% These results indicate that \textbf{\name{}\zhenhui{'s generative images} can improve the long-term learning gains on using target words in their verbal expressions compared to the baseline system}.
}

\qiaoyi{
As for the fluency of participants' spoken English in the immediate posttest (\autoref{fig:learing_outcoming3}), participants can tell the story significantly more fluently after the learning session with \name{} $(M = 6.604, SD = 0.935)$ that that with the baseline system $(M = 6.354, SD = 0.872)$.
Similarly, in the delayed posttest, participants' spoken English in telling the story with target words is significantly more fluent after learning with \name{} $(M = 6.229, SD = 0.901)$ compared to the baseline system $(M = 5.688, SD = 0.966)$.
The results of repeated measures ANOVA indicate that both the system factor (\name{} and Baseline) and the time factor (immediate posttest vs. delayed posttest) significantly affect the fluency of participants' spoken English (system factor: $p < 0.05, F = 4.01, \eta^2=0.041$; time factor: $p < 0.001, F = 21.552, \eta^2=0.238$). 
}
% However, as for fluency in expression, participants perform significantly better utilizing \name{} $(M = 6.604, SD = 0.935;\; p<0.05)$ than using the baseline system $(M = 6.354, SD = 0.872)$. 
% In the delayed posttest, With \name{}, participants demonstrate significant advantages in the number of target words used $(p<0.05)$, the number of target words pronounced correctly $(p<0.1)$, and the number of target words used correctly $(p<0.01)$. 
% Furthermore, participants utilizing \name{} $(M = 6.229, SD = 0.901;\; p<0.001)$ exhibit notably greater expressive fluency in comparison to those using the baseline system $(M = 5.688, SD = 0.966)$.
% ``I find myself expressing much more smoothly when using \name{}. The images aid my memory, allowing me to recall specific details and the following sentences of the story scene as I retell the story.'' (P10)
These results suggest that \textbf{\name{}\zhenhui{'s generative images} can significantly improve the learners' fluency in using target words to tell a story after the story retelling practices compared to the baseline system}. 
% \textbf{These results suggest that \name{} can improve learning gains on pronouncing the target words and using them correctly and fluently in the story contexts after a one-week delay.}
}

\begin{figure}[htbp]
    \centering
    \includegraphics[width=14cm]{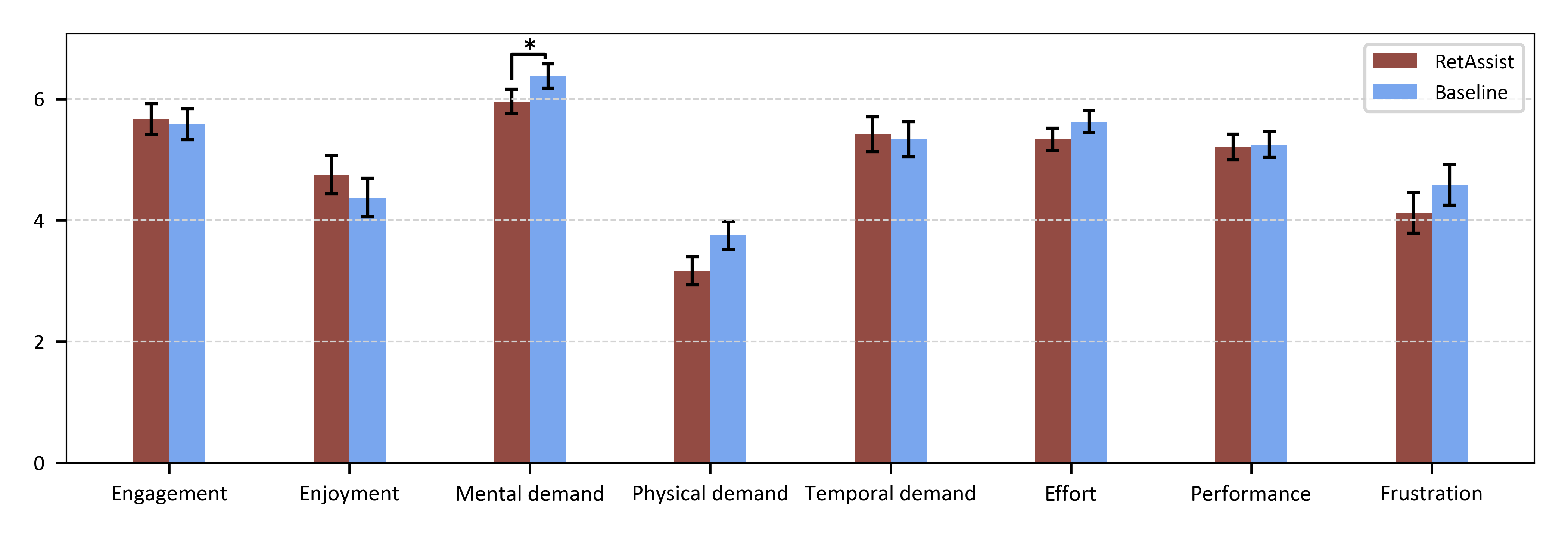}
    \Description[Bar graph]{Bar graph indicating a significant difference between RetAssist and Baseline in Mental demand. }
    \vspace{-0.5cm}
    \caption{
     RQ2 results regarding engagement, enjoyment, and workload in vocabulary learning sessions. \qiaoyi{$*:p<0.05$ using paired samples Wilcoxon signed rank tests. }
    }\label{learning_exp_e_6}
\end{figure}
\vspace{-0.2cm}

\begin{figure*}[htbp]
    \centering
    \begin{minipage}[t]{0.32\textwidth}
        \centering
        \includegraphics[width=5cm]{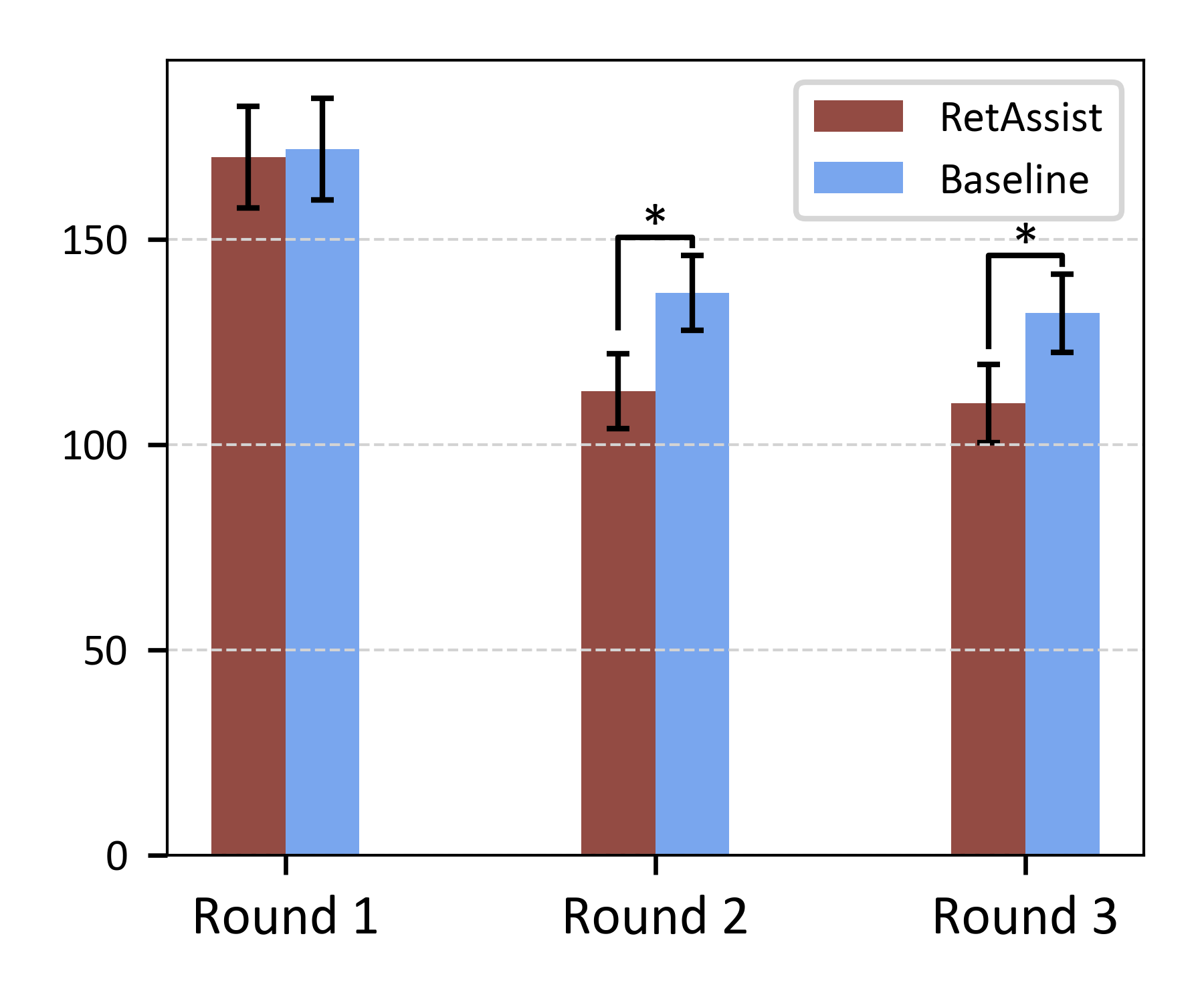}
        \Description[Bar graph]{Bar graph indicating significant differences between RetAssist and Baseline in Round 2 and Round 3. }
        \vspace{-0.7cm}
        \caption{RQ2 results regarding time spent by users in three rounds of retelling. \qiaoyi{$*: p<0.05$ using paired-sample t-tests. }}
        \label{time_e_3}
    \end{minipage}%
    \hspace{0.2cm}
    \begin{minipage}[t]{0.32\textwidth}
        \centering
        \includegraphics[width=5cm]{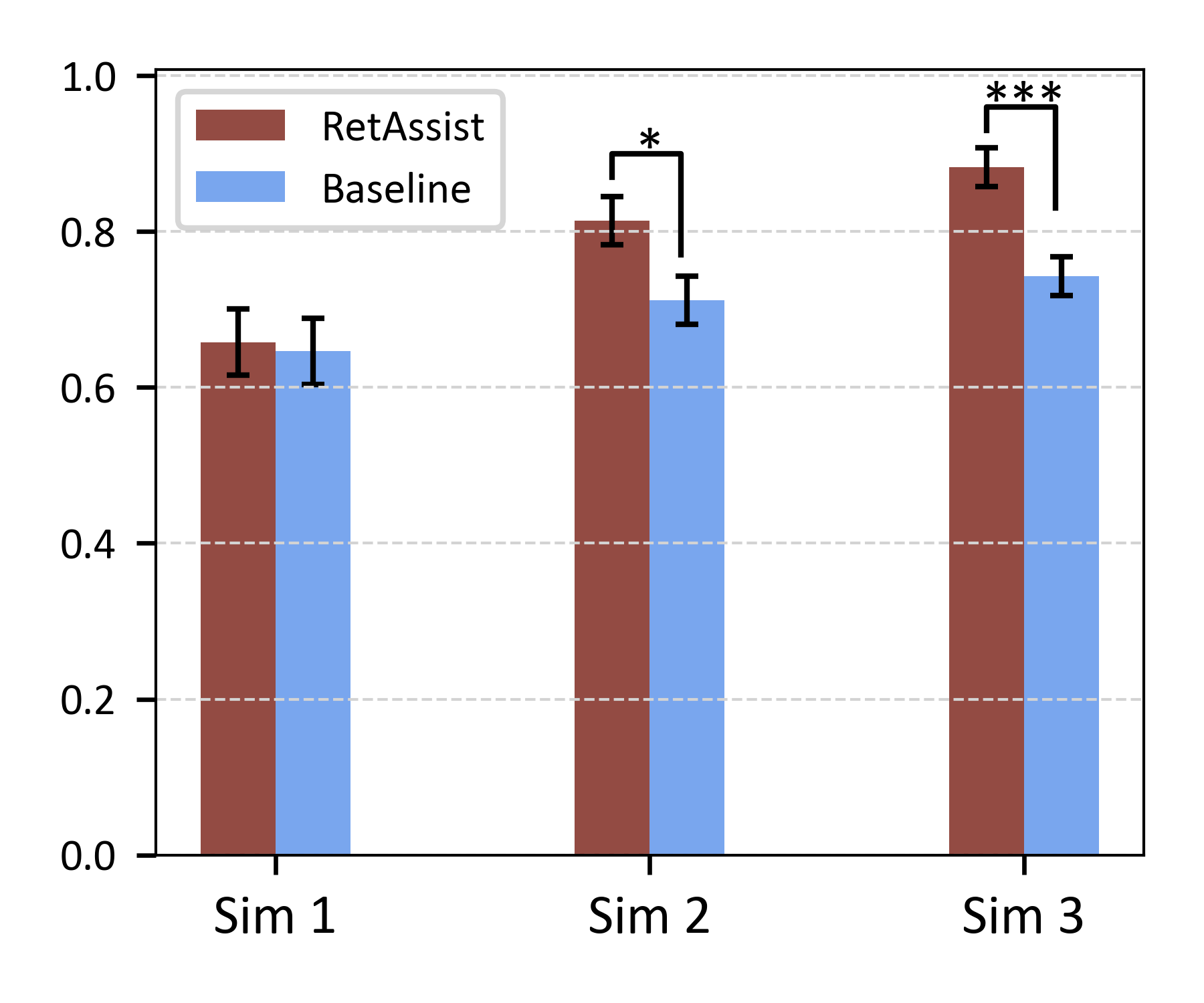}
        \Description[Bar graph]{Bar graph indicating significant differences between RetAssist and Baseline in Sim 2 and Sim 3. }
        \vspace{-0.7cm}
        \caption{
    RQ2 results regarding the semantic similarity between users' retelling content and story.  \qiaoyi{$*: p<0.05, ***: p<0.001$ using paired-sample t-tests. }
    }\label{fig:sim_e_3}
    \end{minipage}%
    \hspace{0.2cm}
    \begin{minipage}[t]{0.32\textwidth}
        \centering
        \includegraphics[width=5cm]{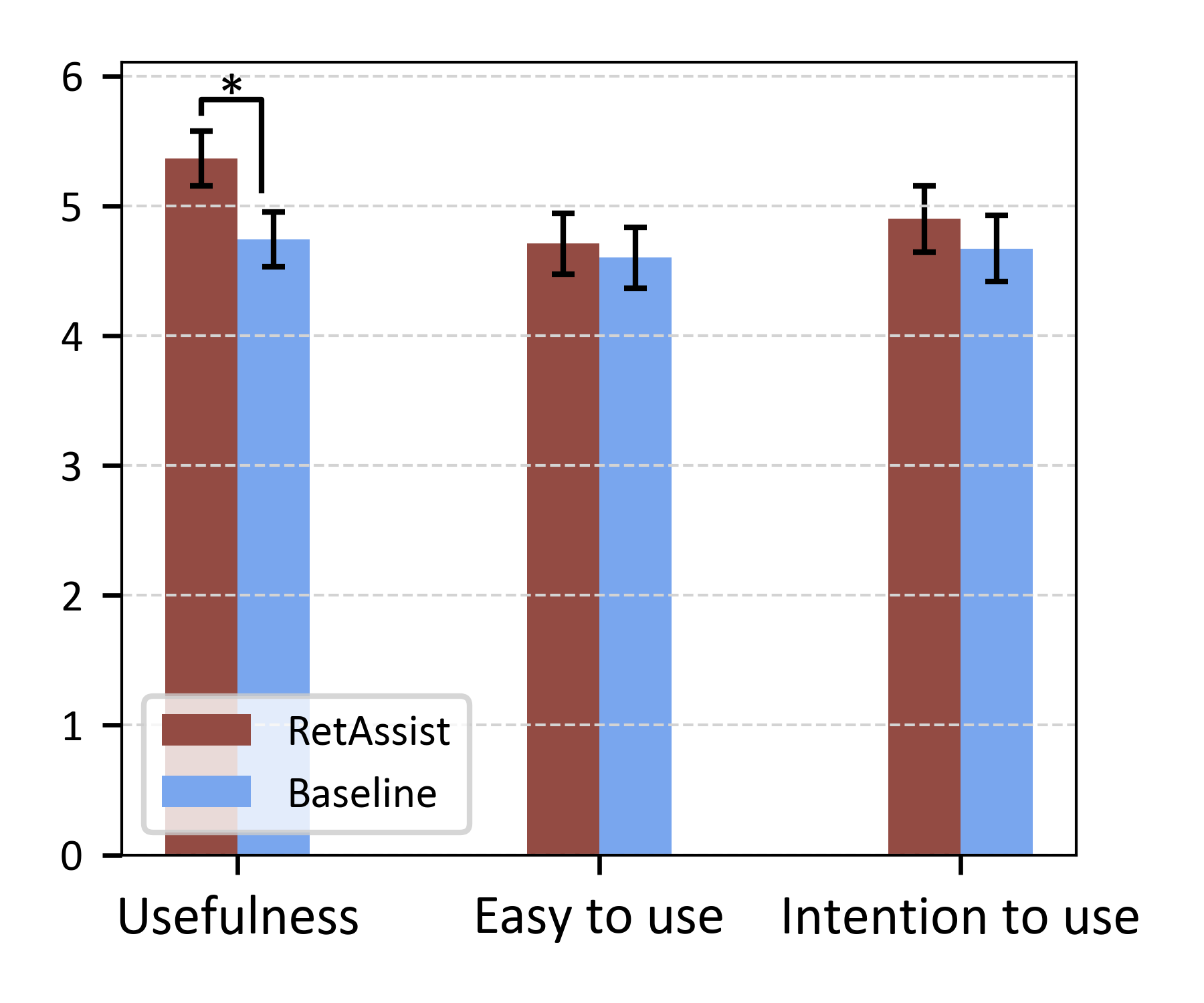}
        \Description[Bar graph]{Bar graph indicating a significant difference between RetAssist and Baseline in Usefulness. }
        \vspace{-0.7cm}
        \caption{
    RQ3 results regarding user perceptions of each
interface. \qiaoyi{$*: p<0.05$ using paired samples Wilcoxon signed rank tests. }
    }\label{use_perceptions}
    \end{minipage}%
\end{figure*}

\peng{
\subsection{Learning Process (RQ2)}
\subsubsection{Engagement, enjoyment and workload.}
\qiaoyi{
% As shown in \autoref{learning_exp_e_6}, 
% participants generally feel that they are more engaged and enjoyed the vocabulary learning process with \name{} than the baseline system.
As shown in \autoref{learning_exp_e_6}, participants report a slight increase in engagement and enjoyment during the vocabulary learning process with \name{} compared to the baseline system, though the difference was not statistically significant.
% participants mildly but not significantly feel that they are more engaged and enjoyed the vocabulary learning process with \name{} than the baseline system.
% engagement and enjoyment when using A was not significantly better than when using B.
However, \chenqy{22 (out of 24)} participants commend the quality of the images and feel that the images are closely aligned with the text.
``The pictures are appealing, and I can interact with them by switching the picture and checking its related sentence'' (P12).
}
\qiaoyi{
Furthermore, participants report a lower level of mental demand $(p<0.05, z = 2.066, \text{Cohen's d} = 0.455)$ 
% physical demand $(p<0.1)$, and frustration $(p<0.1)$ 
during the vocabulary learning process with \name{} than that with the baseline system. %significantly 
}
\chenqy{21} participants perceive that practicing story retelling with the baseline system is notably more challenging, as they need to mentally visualize and construct the scene of the story. 
% ``With the baseline system, I can only depend on abstract connections formed through text. In contrast, \name{} is much more intuitive and effective at recalling the storyline.'' (P6)
``Recalling the story's details and scenarios (with the baseline system) takes up a lot of my mental effort. In contrast, \name{} helps me to recall the story in a visual way'' (P6). %requires a significant effort
Five participants further report that the baseline system is monotonous compared to \name{}. 
``I do not like the (baseline) interface as it is monotonous and inflexible'' (P3).

% Based on the interviews, we have some interesting observations: 
% a) Participants willingly mentioned that the inclusion of images significantly enhanced their engagement; 
% a) Twenty-two participants commend the quality of the images and feel that the images are closely aligned with the text.
% ``The pictures are appealing, and I can interact with them by switching the picture and checking its related sentence.'' (P12)
% b) Time constraints can lead to heightened participant engagement but could decrease the overall enjoyment of the learning experience; 
% Nine participants commend that the countdown instills a sense of urgency, motivating them to fully commit to their studies.
% ``The countdown really drives me to stay completely engaged, but it also makes me feel anxious and stressed out.'' (P22)
% c) Participants perceive the baseline system as notably more challenging, as they are required to mentally visualize and construct the scene. 
% ``With the baseline system, I can only depend on abstract connections formed through text. In contrast, \name{} is much more intuitive and effective at recalling the storyline.'' (P6)

\subsubsection{Performance in each round of repeated retelling.}
% \textbf{Task complete time and semantic similarity.}
\autoref{time_e_3} shows the spent time in each round of repeated retelling with \name{} and the baseline system.
% Regarding the spent time in each round of repeated retelling with \name{} and the baseline system, as shown in Figure \ref{time_e_3}, there are significant differences between these two systems in the three retelling practices.
\qiaoyi{
Participants spend less time with \name{} in the second $(M (SD): 113 (45.07)$ vs. $137 (61.02); p<0.05, t = -2.227, \text{Cohen's d} = 0.321)$ and third $(110 (47.23)$ vs. $132 (55.28); p<0.05, t = -2.18, \text{Cohen's d} = 0.315)$ rounds of repeated retelling compared to the cases with the baseline system.
}
%$(M: 113/137,\ SD: 45.07/61.02;\; p<0.05)$ and third $(M: 110/132,\ SD:47.23/55.28;\; p<0.05)$ rounds compared to the cases with the baseline system.
``While using the baseline system, I frequently run out of the limited time before finishing retelling the story; however, when using \name{}, I am more comfortable in the repeated retelling stage and can complete the retelling on time'' (P20). %f% I can comfortably complete the task within the time limit'' (P20). %find myself running short on time
\qiaoyi{
Meanwhile, as shown in \autoref{fig:sim_e_3}, the semantic similarity between users' retelling content and original story significantly increases during the second $(p<0.05, t = 2.397, \text{Cohen's d} = 0.346)$ and the third rounds $(p<0.001, t = 3.793, \text{Cohen's d} = 0.547)$ of repeated retelling. 
}
Nineteen participants attribute their improvement during the practice to the provided images in \name{}. 
``Images provided in \name{} help me better connect my native language expression and English expression of the target words, which helps me reflect on the details and storyline in a short time'' (P4). 
% I connected the images provided in \name{} with the story, which helped me reflect on the details and storyline in a short time
% non-verbal modalities such as images bridge the gap between two different languages, which would enhance the likelihood of recalling the second language
\textbf{These results indicate that compared to the baseline system, \name{} can reduce learners' workload and improve their efficiency and performance of each round of repeated retelling during the story retelling practices.} 
% \textbf{These indicate that participants obtain significantly greater progress in the proficiency and quality of retelling with \name{} than the baseline system.} 
}

\peng{
\subsection{Perceptions towards the Systems (RQ3)}   %Measures of User 
% \textbf{Usefulness.} 
As shown in \autoref{use_perceptions}, 
% participants feel significantly different regarding the usefulness of our two systems. Specifically, they 
\qiaoyi{
participants feel that our \name{} $(M = 5.365, SD = 1.031)$ is significantly more useful than the baseline system $(M = 4.74, SD = 0.996;\; p < 0.05, z = 2.29, \text{Cohen's d} = 0.604)$. 
}
Nineteen participants implied that the images in \name{} are the reason for rating it more useful. % practically assist them in their story retelling practices. 
``Without the images, I find it difficult to go through the repeated retelling stage. The images are especially useful when I am stuck'' (P13). 
% P1, P5  and P17 reflected that vocabulary learning process with Read-and-Retell is tedious and stressful. They can easily give up in this way. 
% Especially P13 pointed out, “Without the assistance of pictures and prompts, repetition is difficult, it is equivalent to pure brain activity.”
% Furthermore, a substantial divergence is evident in the perceived utility of the two systems. Nineteen participants consistently reported significantly higher levels of practical efficacy in their learning experiences with \name{} $(M = 5.36, SD = 0.21, p < 0.05)$ than the baseline system $(M = 4.74, SD = 0.203)$.
% \textbf{Easy to use.} 
% As for the second aspect of the technology acceptance model, 
There is no significant difference between \name{} $(M = 4.708, SD = 1.156)$ and the baseline system $(M = 4.604, SD = 1.141)$ regarding easiness of use.
% Participants feel that \name{} $(M = 4.708, SD = 1.156)$ is easier to use than the baseline system $(M = 4.604, SD = 1.141)$.
We have comments from twenty-one participants that praise the interaction design of \name{}. % flexibility in use.
``The interface of \name{} is intuitive, and the interaction flow is clear. I can listen to the story while easily reading the story with aligned images'' (P11).
% ``\name{} meets my story retelling needs by listening to the story audio and checking the translation. Its functions are intuitive and clear.'' (P11)
% ``\name{} meets my vocabulary learning needs. Not only that, it is intuitive and clear, and I can easily know how to use its functions'' (P11).
% \textbf{Intention to use.} 
Lastly, participants generally have a higher intention to use \name{} $(M = 4.9, SD = 1.249)$ for vocabulary learning in the future compared to baseline system $(M = 4.6, SD = 1.337)$. 
% Five participants expected they could use \name{} in their future vocabulary learning activity. 
Twenty participants comment that they prefer the \name{} for future vocabulary learning. % but not the baseline system.
``With \name{}, I can express the learned words more correctly with less pressure. I want to have it as my weekly used vocabulary learning system'' (P16). 
However, four participants prefer the baseline system, because they feel it is time-consuming to view the images and mentally connect them with the story. % the images may cost them extra time to align with the text. 
}

%am willing to use it to learn vocabulary efficiently
% Five participants conveyed that they would be down to use the \name{} system to learn words if it was developed.
% ``By using \name{}, I can remember words more deeply with less pressure. I am willing to use it to learn vocabulary efficiently. '' (P18).

% {?? These statements are more about the usefulness and easy to use. For the quotes about intention to use, you may want to talk about how they can be used in their daily vocabulary learning activities / story retelling practices.}
% }

% \begin{figure*}[htbp]
%     \centering
%     \includegraphics[width=8cm]{figure/impost.png}
%     \caption{Expressive scoring on a 9-point scale for vocabulary learning tasks in the immediate post-test (*: $p$ < .05). }\label{express_im}
% \end{figure*}

% \begin{figure*}[htbp]
%     \centering
%     \includegraphics[width=14cm]{figure/depost.png}
%     \caption{Expressive scoring on a 9-point scale for vocabulary learning tasks in the delayed post-test (*: $p$ < .05). }\label{express_de}
% \end{figure*}

\peng{
\section{discussion}
% \pzh{
% \subsection{Story Retelling for Vocabulary Learning}
In this work, we develop the \name{} system that aims to facilitate vocabulary learners in their story retelling practices. 
\chen{
Its core features are the generative images relevant to the story in the story comprehension and repeated retelling stages.  
% Our proposed computational workflow to generate images related to the story enables individual vocabulary learners to access relevant images to any story content in their retelling practices. 
% [Our within-subjects study (N=24) demonstrates that \name{} significantly enhances the learning outcome on mastering meanings and expressions of target words than the baseline system without our proposed image prompts. For the learning experience, \name{} significantly eases the learning workload and is perceived as more useful. ]
% Our within-subjects user study with 24 participants indicates that learning with \name{} significantly eases the learning workload and is perceived as more useful than the baseline system.  
% As for the learning outcome, 
Our study shows that participants using either \name{} or the baseline system can master the meanings and expressions of the target words right after a story retelling practice, supporting that story retelling is an effective approach to vocabulary learning \cite{dunst2012children, morrow1985retelling, merritt1989narrative, gibson2003power}. % (\autoref{fig:learing_outcoming1})
However, one week after the practices, participants better recall and verbally express the target words learned with \name{} than those with the baseline system. 
This proves the value of our generative images for supporting vocabulary learning and provides empirical evidence for the benefits of visual aids for language learning stated in the Cognitive Theory of Multimedia Learning \cite{paivio1980dual}. 
%, indicating that \name{} further lead to more learning gains on retention of words' meanings (\autoref{fig:learing_outcoming1}) and using target words in verbal expressions (\autoref{fig:learing_outcoming2} and \autoref{fig:learing_outcoming3}), especially on the correct contextualized usage and pronunciations. 
% This finding empirically supports the theory of scaffolding \cite{vygotsky1978mind}. 
% As suggested by our participants, the images help \wxb{understand and recall the vocabulary meaning and its usage. 
% % to understand and recall. 
% This is the image benefits implied by
% % inclined with the benefits of images implied by 
% the Cognitive Theory of Multimedia Learning \cite{paivio1980dual}.}
% As suggested by our participants, images related to the story help them connect English with their native language, which assists their understanding and recall of vocabulary meaning and its usage. 
% These are the benefits of visual aids implied by the Cognitive Theory of Multimedia Learning \cite{paivio1980dual}. 
% The images contribute to this improvement, as suggested by the Cognitive Theory of Multimedia Learning \cite{paivio1980dual}. 
}

\subsection{Design Considerations} %  for Vocabulary Learning Support Tools
% \pzh{
% The design and evaluation of \name{} 
Based on our findings, we provide three design considerations for story-based vocabulary learning tools. % point out two important directions for further development of computer-assisted vocabulary learning tools:

% \textbf{\textit{Adaptive story visualization strategies. }}
\textbf{\textit{Provide more types of visual aids. }}
% \textbf{\textit{Visualize the learning materials. }}
Participants generally favor \name{}'s images for helping them comprehend and recall stories. 
% \name{} offers images relevant to the story to assist users in making connections between each image and sentence, thus supporting their story comprehension and recollection \cite{oktarina2020effectiveness, filippatou1996pictures, paivio1980dual}.
% Up to fifteen participants consider that these images attached to the story assist them in making connections between each image and sentence, thus supporting their story comprehension and recollection \cite{oktarina2020effectiveness, filippatou1996pictures, paivio1980dual}. 
However, three participants comment that they still have difficulty in recalling the expression of target words with the generative images and suggest that it would be better to visualize stories through mind maps or flow charts \cite{praneetponkrang2014use, musti2022virtual}. 
% However, learners who are not sensitive to image modality often have difficulty making this connection. 
% Three participants hoped to obtain other visual aids to support them in sorting out and recalling the story framework, like a flowchart of the story's development \cite{praneetponkrang2014use, musti2022virtual}. 
Moreover, participants expect \name{} to incorporate short videos \cite{bal2014investigation} or motion graphics \cite{larsari2020psychological} into vocabulary learning.
``Understanding the images themselves is an additional burden for me. I would prefer a more explainable form of visual aids to help me understand some abstract storylines in the story comprehension stage'' (P18).
We, therefore, suggest that the generative technique could offer other forms of visual aids such as an extracted mind map and a relevant video clip, and allow users to customize them based on their interests. 
% ``Understanding the images themselves is a burden for me. I would prefer a more interpretable, text-related means of story visualization to aid my story comprehension and recall.'' (P12)
% It reflects that visual aids offered in the learning system could accommodate learners with different levels of image sensitivity with different styles of visual aids. 
% To enable users to make better use of the visual aids provided, the learning tools can conduct interpretable story visualization that explicitly attaches some interpretable labels to the generated story-related images in order to construct them into a mind map format.
% }
% Therefore, the learning tool will provide images relevant to the story for users who are sensitive to image modality. 
% In contrast, the learning tools summarize the flow chart of the story development.

% \pzh{
\textbf{\textit{Offer prompts that are adaptive to users' performance. }}
\name{} currently provides fixed image prompts and word prompts. 
However, two participants suggest that they need more personalized and interactive prompts. 
For example, the system can ``recognize my stuckness and give me corresponding hints based on the progress of my current retelling.'' 
To provide timely and personalized support during each round of repeated retelling, it would require future researchers to label a set of story retelling audio clips for training a model to predict users' difficult timing based on their tone, speed, and pauses in the current practice. %,  thus providing more interactive and personalized services based on the user's performance. 

\textbf{\textit{Provide suggestions to improve.}}
% \name{}'s feedback after each round of the retelling exercise is more like an assessment of how well users were.
\wxb{The feedback offered by \name{} includes the correctness of \chen{target words' semantic usage as well as highlighting} the incorrectly used target words and the corresponding sentences. }
Four participants expect that it can also explicitly tell them how to improve in the next round of practice. 
For instance, the system can ``correct mispronunciations of words, list the target word's grammatical usage and provide additional example sentences'' to enhance the comprehension of the target vocabulary. 
We suggest that future vocabulary learning tools should offer not only feedback on what and why a target word is misused but also suggestions on how to deepen the understanding of this word, \eg with more example sentences. % generative models can provide extended learning materials on the misused target words in the feedback. 

% For example, in the feedback panel (\autoref{fig:2} C), it can leverage text generation models to offer other example sentences using the misused words in a similar context to the retold story. 
% The improvement feedback offered by \name{} is more like a collation of key learning materials from the original story based on the user's retelling performance.
% Although it effectively enables learners to review their unfamiliar parts better, it lacks some extended learning.
% Therefore, future vocabulary learning tools could consider feedback with more relevant learning materials targeting the user's weaknesses such as example sentences that are similar to the context of the misused target word in the story. 
% It is also a good idea to predict the most likely error and stuck parts in the next retelling from the performance of the previous time so that relevant learning materials can be provided before this feedback to optimize the user's next retelling performance.
% }

\subsection{Broader Impact to Generative AIs for Education}
% 由我们的design引申出的generative content as the learning materials的positive potentials，generative content 作为learning materials要注意的地方（negatives，比如生成harmful content，喧宾夺主（没有让用户付出足够的learning effort (可以cite storyfier那篇文章的情况)）等，需要像我们一样让teachers和learners参与到设计过程中mitigate potential negative impact），我们的framework和我们的系统更务实的教育应用（比如小孩子的故事书绘图，应用到线下/线上实际教学的可能性和要注意的地方等等）
\cqyrevise{
% [由我们的design引申出的generative content as the learning materials的positive potentials]
Our design and development of \name{} offers a feasible example of leveraging generative AIs to support learning tasks. 
First, generative models can offer meaningful and flexible learning materials, \eg ChatGPT \cite{bang2023multitask} that generates a story given any target words in our case. It is also promising to apply these models to prepare listening materials \cite{ren2019fastspeech} and provide contextually personalized learning materials \cite{draxler2023relevance} for language learners. 
Second, generative models can support additional modalities of learning activities used in traditional instruction on a large scale.
% In addition to assisting memory, text-to-image AI can be integrated into 3D Design Workflows to produce reference images, prevent design fixation, and inspire design considerations \cite{liu20233dall}. 
In addition to serving as visual aids as in our case, text-to-image AI can be integrated into 3D Design Workflow to produce reference images, prevent design fixation, and inspire design considerations \cite{liu20233dall}. 
Also, they can empower a conversational agent, which acts like a lecturer, to socially converse with the learners to practice their spoken language \cite{ruan2021englishbot} on any topic.
% [generative content 作为learning materials要注意的地方（negatives，比如生成harmful content，喧宾夺主（没有让用户付出足够的learning effort (可以cite storyfier那篇文章的情况)）等，需要像我们一样让teachers和learners参与到设计过程中mitigate potential negative impact]

However, utilizing generative content as learning materials may have the potential to hinder learning gains in certain scenarios.
One concern is the risk of generative AIs in terms of accuracy and reliability. Learners need to take precautions against generating errors or false information when adopting generated content as learning material, and the generative content may be one-sided and outdated because of the limitations of the training data for generative AIs \cite{educsci13040410}.
Another \zhenhui{concern} is that \zhenhui{the assistance of generative AIs may discourage users from putting in enough effort in the learning process.} %learning process assisted by generative AIs does not allow the learner to put in enough effort. %priority 
For instance, Peng \etal's study suggests that learners \zhenhui{could} experience \zhenhui{reduced} gains in vocabulary acquisition when engaged in writing exercises with \zhenhui{generative AIs} compared to those without AI assistance \cite{peng2023storyfier}. %diminished AI co-writing
This could be attributed to the fact that participants invested less time in writing and wrote significantly fewer words in the story, as it indicates a preference for dependence on the generative model for assistance \cite{peng2023storyfier}. 
To mitigate \zhenhui{these} potential negative impacts, \zhenhui{we suggest that the developers of learning support systems should examine the quality of generated content beforehand and work with targeted learners and educators to identify proper design principles (\autoref{process}). }
% our study provides a strategy of involving educators and educated in the design process to explore the design principles of the learning supporting system (\autoref{process}).

% System generalizability and ethical implications
% Beyond the independent utilization of \name{} by learners outside of the classroom, 
\xingbo{Although \name{} is initially designed for independent learning outside of the classroom, it can be useful in diverse educational settings beyond individual study.}
% we also believe that 
For example, teachers in traditional classrooms can use \name{} to enrich vocabulary instruction around story reading or story retelling. 
In addition to assisting ESL learners in vocabulary acquisition using story retelling, the system's story text-to-image generation workflow is expected to be useful in general education scenarios that combine images with stories. 
For example, our workflow can generate sentence-level illustrations for children's storybooks to help them better understand the meaning of textual descriptions. 
In addition, for cultivating children's expressive language skills, our story text-to-image generation workflow can be used as an interactive and creative way for teachers or parents to practice expressive language in children's education. 
By retelling stories with their illustrations, children can develop the ability to clearly organize their verbal expressions, make associations between visual materials and textual materials, and creatively conceptualize the plot with the illustration details. 
Despite the enormous potential of our proposed system and workflow in education, we must approach potential risks cautiously to ensure that they bring positive and sustainable impacts. 
We must ensure that the generative images and stories conform to widely accepted educational standards and ethical norms to avoid conveying incorrect information or inappropriate content. 
}

\subsection{Limitations and Future Work}
% \pzh{
Our study has several limitations that urge future work. 
First, as our primary focus is on vocabulary learning support, we did not examine \name{}'s impact on learners' story retelling skills. 
Learners may have overreliance on generative images
% rely on  the generative images a lot 
in the story retelling practices, while in the English exams that test story retelling performance, they would not have such assistance. 
Future work can extend \name{} for training story retelling skills. 
Second, we evaluate \name{} with twenty-four English-as-second-language undergraduates learning IELTS words, who could not represent all target user groups. 
\chenqy{
% We encourage future researchers to customize our system and evaluate it to support vocabulary learners of different ages, expertise, and cultures (\eg middle or high-school students and English students learning Chinese). 
We would like to extend the study to include learners of different age groups or proficiency levels in our future work, and we also encourage future researchers to customize our system and evaluate it to support vocabulary learners of different ages, expertise, and cultures (e.g., middle or high-school students and English students learning Chinese). 
}
Third, we conducted a short-term user study that can reveal \name{}'s user experience and effectiveness \chen{in our proposed learning tasks. }    % a controlled lab environment. 
To examine its usage in the wild, we need a long-term field study in which users can specify any target words and take story retelling practices at any time they want. 
Fourth, we design our computational workflow of generating multiple image prompts relevant to each story. \wxb{In our formative study, we indicate that the stories are short ones with approximately 60 words. }
% the proper length of the story here is around
% a short story of about 60 words. 
However, this study design may not apply to all user groups. 
As the story gets longer, our computational workflow will generate more sentence-level images that may decrease the coherence among images and increase users' cognitive workload to process them in the story retelling practices. 
To alleviate this cognitive load, future work could consider ways to generate images based on the semantic segments of the story (\ie one or multiple sentences that describe one image). % alternative ways of dividing stories or alternative means of story visualization. 
\chenqy{
Fifth, future design iterations of \name{} could incorporate more advanced AI features like adaptive learning algorithms that tailor image selection or presentation based on individual learner performance.
Sixth, our study exclusively utilized generative images as visual aids, yet alternative media formats might yield different outcomes. 
% we provide generative images as the only type of visual aid in our study, but other types of visual aids could impact differently. 
In order to understand the affordances of static or dynamic images for learning, we will consider comparing the efficacy of generative images with other media types (\eg videos or interactive graphics) in our future work.
}

\section{conclusion}
\chen{
In this paper, based on educational literature and working with teachers as well as ESL learners, we iteratively design and develop an interactive system, \name{}, to facilitate vocabulary learners in story retelling practices. % story-retelling-based vocabulary learning. 
\name{} equips our proposed computational workflow that generates images relevant to the story to foster users' understanding and recall of the story that contains a set of target words. % and their contextualized usage. 
We conduct a within-subjects study with 24 participants in comparison to the baseline system without generative images. 
Our results show that learning with \name{} leads to significantly better learning outcomes on mastering meanings and expressions of target words than learning with the baseline system. 
Our work demonstrates the feasibility and effectiveness of generative models to support language learning tasks and offers implications for future learning support tools.
% For the learning experience, \name{} significantly eases the learning workload and is perceived as more useful. 
}
}

\begin{acks}
This work is supported by the Young Scientists Fund of the National Natural Science Foundation of China (NSFC) with Grant No.: 62202509, NSFC Grant No.: U22B2060, and the General Projects Fund of the Natural Science Foundation of Guangdong Province in China with Grant No.
2024A1515012226. 
Also, this work is supported in part by HKUST 30 for 30 with Grant No.: 3030\_003.
We are grateful to the anonymous reviewers for their insightful suggestions. 
\end{acks}

\bibliographystyle{ACM-Reference-Format}
\bibliography{main}

\end{document}